\documentclass[9pt,twocolumn,twoside]{pnas-new}

\usepackage{color}
\usepackage{mathtools}
\usepackage{amsmath}
\usepackage{amssymb}
\usepackage{graphicx}
\usepackage{hyperref}

\templatetype{pnasresearcharticle} 

\title{Topological transition in measurement-induced geometric phases}

\author[a,b,1]{Valentin Gebhart}
\author[b,1]{Kyrylo Snizhko}
\author[a]{Thomas Wellens}
\author[a]{Andreas Buchleitner}
\author[c]{Alessandro Romito}
\author[b,2]{Yuval Gefen}

\affil[a]{Physikalisches Institut, Albert-Ludwigs-Universität Freiburg, Hermann-Herder-Str.~3, 79104 Freiburg, Federal Republic of Germany}
\affil[b]{Department of Condensed Matter Physics, Weizmann Institute of Science, Rehovot 76100, Israel}
\affil[c]{Department of Physics, Lancaster University, Lancaster LA1 4YB, United Kingdom}

\leadauthor{Gebhart}

\significancestatement{We bring together two concepts at the forefront of current research:
	measurement-induced back-action (resulting in the steering of a quantum
	state), and topological transitions. It has been widely believed
	that the transition from the limit of a strong (projective) measurement
	to that of weak measurement involves a smooth crossover as function
	of the measurement strength. Here we find that varying the measurement
	strength continuously may involve a topological transition.
	Specifically, we address measurement-induced geometric phase, which
	emerges following a cyclic sequence of measurements. Our analysis
	suggests that measurement-based control and manipulations of quantum
	states (a promising class of quantum information protocols) may involve
	subtle topological features that have not been previously appreciated.}

\authordeclaration{The authors declare no conflict of interest.}
\equalauthors{\textsuperscript{1}V.G. and K.S. contributed equally to this work.}
\correspondingauthor{\textsuperscript{2}E-mail: yuval.gefen@weizmann.ac.il}


\begin{abstract}
The state of a quantum system, adiabatically driven in a cycle, may
acquire a measurable phase depending only on the closed trajectory
in parameter space. Such geometric phases are ubiquitous, and also
underline the physics of robust topological phenomena
such as the quantum Hall effect. Equivalently, a geometric phase may
be induced through a cyclic sequence of quantum measurements. We show
that the application of a sequence of
weak measurements renders the closed trajectories, hence
the geometric phase, stochastic. We study the concomitant probability
distribution and show that, when varying
the measurement strength, the mapping between the measurement sequence
and the geometric phase undergoes a topological transition. Our finding may impact measurement-induced control and manipulation
	of quantum states---a promising approach to quantum information processing.
	It also has repercussions on understanding the foundations of quantum
	measurement.
\end{abstract}

\dates{This manuscript was compiled on \today}
\doi{\url{www.pnas.org/cgi/doi/10.1073/pnas.XXXXXXXXXX}}

\begin{document}
	\global\long\def\sgn{\mathrm{sgn}}%
	\global\long\def\ket#1{\left|#1\right\rangle }%
	\global\long\def\bra#1{\left\langle #1\right|}%
	\global\long\def\sp#1#2{\langle#1|#2\rangle}%
	\global\long\def\abs#1{\left|#1\right|}%

	\maketitle
	\thispagestyle{firststyle}
	\ifthenelse{\boolean{shortarticle}}{\ifthenelse{\boolean{singlecolumn}}{\abscontentformatted}{\abscontent}}{}
	
	\dropcap{T}he overall phase of a system's quantum state is an unmeasurable quantity
	that can be freely assigned. However, when the system is driven slowly
	in a cycle, it undergoes an adiabatic evolution which may bring its
	final state back to the initial one \cite{Born1928,Messiah1961}; the
	\emph{accumulated} phase then becomes gauge invariant and, therefore,
	detectable. As noted by Berry \cite{Berry1984a}, this is a geometric
	phase (GP) in the sense that it depends on features of the closed
	trajectory in parameter space, and not on the dynamics of the process.
	Geometric phases are key to understanding numerous physical effects
	\cite{Wilczek1989,Xiao2010,Chruscinski2012}, enabling the identification
	of topological invariants for quantum Hall phases \cite{Thouless1982},
	topological insulators and superconductors \cite{Bernevig2013,Asboth2016},
	defining fractional statistics anyonic quasiparticles \cite{Wilczek1990,Law2006},
	and opening up applications to quantum information processing \cite{Zhu2005,Nayak2008a}.
	
	Geometric phases are not necessarily a consequence of adiabatic time
	evolution. For any pair of states $\left|\psi_{l}\right\rangle $,$\left|\psi_{m}\right\rangle $
	in Hilbert space, it is possible to define a relative phase, $\chi_{l,m}\equiv\arg\left[\langle\psi_{l}\vert\psi_{m}\rangle\right]$.
	For a sequence of states \cite{Aharonov1987,Pancharatnam1956} $\left|\psi_{k}\right\rangle $,
	$k=0,\dots,N$, for which $\vert\psi_{N}\rangle\propto\vert\psi_{0}\rangle$,
	one can define the total phase accumulated through the sequence (the
	Pancharatam phase \cite{Aharonov1987,Pancharatnam1956})
	\begin{align}
	\chi_{\mathrm{geom}}^{(P)}&=\sum_{k=0}^{N-1}\chi_{k+1,k}\notag \\
	&=\arg\left[\bra{\psi_{0}}\mathcal{P}_{N}\dots\mathcal{P}_{2}\mathcal{P}_{1}\ket{\psi_{0}}\right]=\arg\langle\psi_{0}|\tilde{\psi}_{N}\rangle,\label{definitionphase}
	\end{align}
	where $\vert\tilde{\psi}_{k}\rangle=\mathcal{P}_{k}\dots\mathcal{P}_{2}\mathcal{P}_{1}\vert\psi_{0}\rangle$
	and $\mathcal{P}_{k}=\vert\psi_{k}\rangle\langle\psi_{k}\vert$ is
	the projector onto the $k$-th state. Note that $\vert\tilde{\psi}_{k}\rangle\propto\vert\psi_{k}\rangle$
	is not normalized, which however does not undermine the definition
	of the phase (unless some $\vert\tilde{\psi}_{k}\rangle=0$). Note
	also that $\chi_{\mathrm{geom}}^{(P)}$ is independent of the gauge
	choice of phases of all $\vert\psi_{k}\rangle$. For a quasicontinuous
	sequence of states $\left\{ \vert\psi_{k}\rangle\right\} $, the Pancharatnam
	phase trivially coincides with the Berry phase under the corresponding
	continuous state evolution. Moreover, for \emph{any} sequence $\left\{ \vert\psi_{k}\rangle\right\} $,
	the Pancharatnam phase is equal to the Berry phase associated with
	the trajectory that connects the states $\vert\psi_{k}\rangle$ by
	the shortest geodesics in Hilbert space \cite{Chruscinski2012,Samuel1988}.
	
	The Pancharatam phase can quite naturally be interpreted as a result
	of a sequence of strong (projective) measurements acting on the system
	and yielding specific measurement readouts \cite{Facchi1999}. This
	interpretation is valid for optical experiments observing the Pancharatnam
	phase induced with sequences of polarizers \cite{Berry1996}. Such
	a phase can be consistently defined despite the fact that measurements
	(typically considered an incoherent process) are involved. A generic
	sequence of measurements is an inherently stochastic process. One
	thus expects a distribution of measurement-induced geometric phases,
	determined by the sequences of measurement readouts associated with
	the corresponding probabilities. For a quasicontinuous sequence of
	strong measurements ($N\rightarrow\infty$ and $\lVert\vert\psi_{k+1}\rangle-\vert\psi_{k}\rangle\rVert=O(1/N)$),
	the induced evolution is fully deterministic due to the dynamical
	quantum Zeno effect \cite{Facchi1999}, thus yielding a unique Pancharatnam-Berry
	phase.
	
	Recently, geometric phases induced by \emph{weak measurements} (that
		do not entirely collapse the system onto an eigenstate of the measured
		observable \cite{Aharonov1988}) have been experimentally observed
		\cite{Cho2019a}. The setup of Ref.~\cite{Cho2019a} ensured that
		only one possible readout sequence contributed to the observed phase.
		When considering all possible readout sequences, the system dynamics
		remains stochastic even for quasicontinuous sequences of weak measurements
		\cite{Jacobs2014,Wiseman2010}. In a wider context, employing weak measurements enables
	continuous monitoring of the system through weak measurements, which
	has been successfully employed experimentally for dynamically controlling
	quantum states \cite{Murch2013a,Weber2014,Naghiloo2019,Minev2019}.
	
	In the present study, we define and investigate the geometric phase
	accrued by a quantum state of a two-level system following a sequence
	of measurements and detector readouts with tunable measurement strength.
	We compute the full distribution function of the induced geometric
	phases and analyze the phase of specific trajectories which can be
	singled out by postselecting specific readout sequences. As opposed
	to the case of projective measurements \cite{Berry1996,Facchi1999},
	where the state trajectory (and the resulting phase) are solely determined
	by the measurement sequence and the measurement readouts, the trajectories
	(and the phases) here depend crucially on the measurement strength.
	We mainly focus on the scenarios of a single postselected state trajectory
	and consider the effect of averaging over all possible trajectories
	in the Supplementary Material.
	
	We discover that a topological transition
	vis-à-vis the geometric phase takes place as a function of the measurement
	strength. The transition is topological in the sense that it is related
	to a discontinuous jump of an integer-valued topological invariant.
	Specifically, we consider a family of measurement sequences: The state
	trajectories induced by these sequences form a surface which covers
	a certain area on the Bloch sphere, and our topological transition
	is manifest through a jump of the Chern number associated with this
	surface. Finally, we propose concrete interferometry protocols, which
	allow us to consistently define geometric phases in the presence of
	detectors, and facilitate their detection.

	\section*{Defining geometric phases from variable strength quantum measurements}
	
	\begin{figure}
	\centering
	\includegraphics[width=0.45\columnwidth]{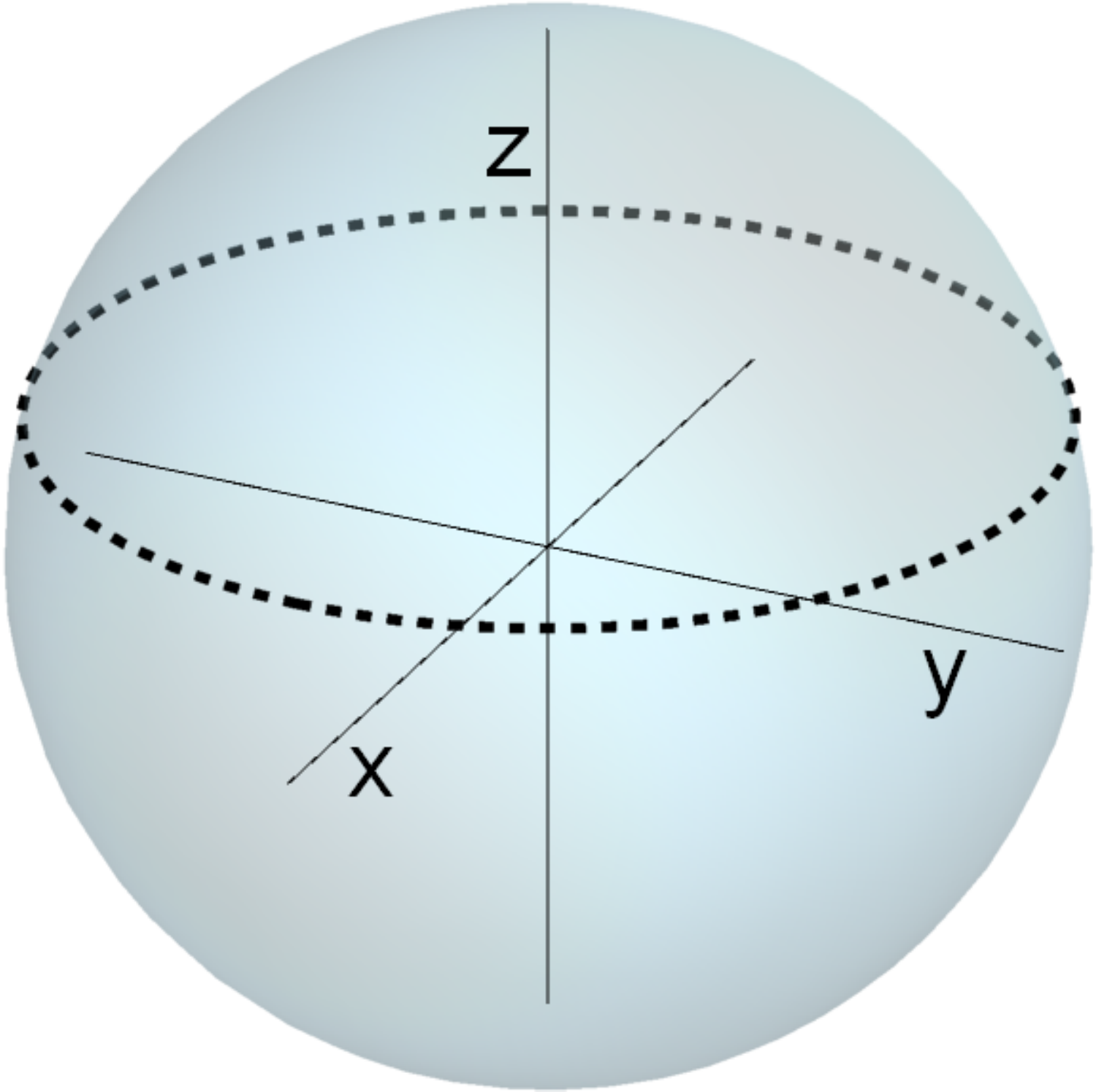} \includegraphics[width=0.45\columnwidth]{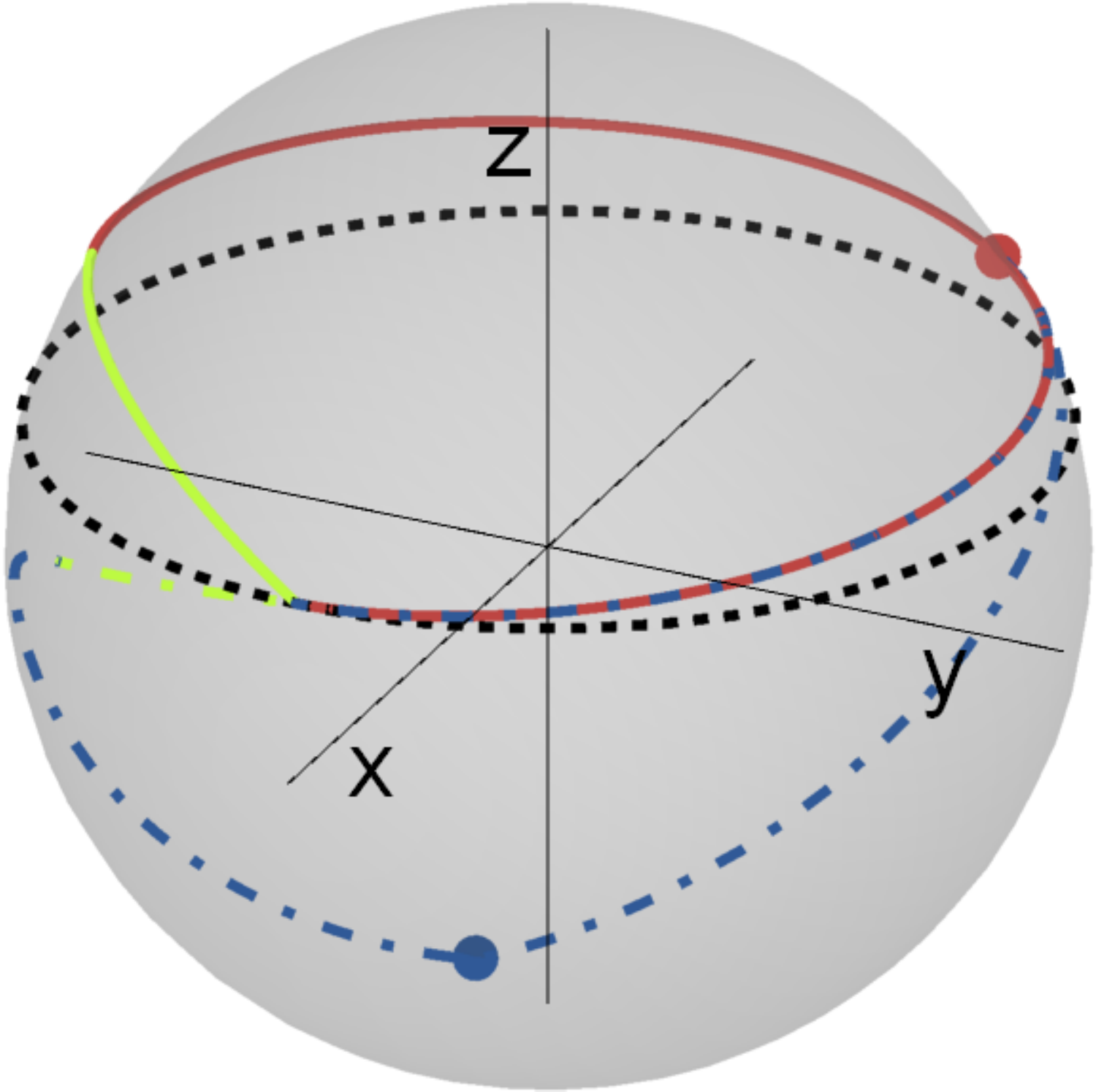}
\caption{\textbf{Measurement sequences
		and induced quantum trajectories.} (left) A measurement sequence spanned
	by the directions $\mathbf{\mathbf{n}}_{k}=(x_{k},y_{k},z_{k})=(\sin\theta_{k}\cos\varphi_{k},\sin\theta_{k}\sin\varphi_{k},\cos\theta_{k})$,
	following a parallel ($\theta_{k}=\pi/4$ and $\varphi_{k}=2\pi k/N$).
	(right) Quantum trajectories on the Bloch sphere induced by the measurement
	sequence depicted on the left for different measurement strengths
	and readout sequences. The trajectory induced by the $\{r_{k}=+\}$
	readout sequence for strong measurements ($c=+\infty$, dotted black)
	meticulously follows the measurement eigenstates $\vert+\mathbf{n}_{k}\rangle$,
	while the $\{r_{k}=+\}$ trajectory for finite-strength measurements
	($c=3$, red) deviates from this line. A weak-measurement-induced
	trajectory ($c=3$) corresponding to the readout sequence with all
	$r_{k}=+$ except for a single $r_{k_{0}}=-$ readout is depicted
	by the dot-dashed (blue) line. The ``$-$'' readout induces a state
	jump from its position on the red line (red dot) to the blue dot via
	the shortest geodesic on the Bloch sphere. For weak measurements,
	the quantum trajectory does not coincide with the measurement sequence
	and may not terminate at the initial state (red and blue lines). The
	final measurement (yellow segment) projects the $N$-th state onto
	the initial state via the shortest geodesic.}
	\label{fig:examples_of_trajectories}
	\end{figure}

	Our system is a qubit whose Hilbert space is spanned by $\ket{\uparrow},\ket{\downarrow}$.
	The system undergoes a chronological sequence of weak measurements,
	labeled as $k=1,\dots,N$. The measured observables are represented
	by the operators $\sigma{}_{\mathbf{n}_{k}}=\boldsymbol{\sigma}\cdot\mathbf{n}_{k}$,
	where $\boldsymbol{\sigma}=(\sigma_{x},\sigma_{y},\sigma_{z})$ is
	the vector of Pauli matrices and $\mathbf{n}_{k}=(\sin\theta_{k}\cos\varphi_{k},\sin\theta_{k}\sin\varphi_{k},\cos\theta_{k})$.
	The sequence of measurement orientations $\left\{ \mathbf{n}_{k}\right\} $
	defines a trajectory on the unit sphere $S^{2}$ as depicted in Fig.~\ref{fig:examples_of_trajectories}
	(left). Each weak measurement is characterized by its strength $\eta\in[0,1]$
	and two possible readouts $r_{k}=+,-$. The modification of the system
	state conditional to the obtained measurement readout $r_{k}$ is
	given by $\vert\psi\rangle\to\mathcal{M}_{k}^{(r_{k})}\vert\psi\rangle$,
	where $\mathcal{M}_{k}^{(r_{k})}=M_{\eta_{k}}(\mathbf{n}_{k},r_{k})$
	are Kraus operators \cite{Wiseman2010,Jacobs2014} (cf. Methods below).
	Given the measurement strength $\eta$, a sequence $\left\{ \mathbf{n}_{k}\right\} $
	of measurement orientations with corresponding readouts $\left\{ r_{k}\right\} $
	induces a sequence of states $\left\{ \vert\tilde{\psi}_{k}\rangle\right\} $
	in the system Hilbert space, where
	
	\begin{equation}
	\vert\tilde{\psi}_{k}\rangle=\vert\tilde{\psi}_{r_{1},\dots,r_{k}}\rangle=\mathcal{M}_{k}^{(r_{k})}\dots\mathcal{M}_{1}^{(r_{1})}\left|\psi_{0}\right\rangle .\label{eq:psif}
	\end{equation}
	The weak measurement-induced geometric phase can be defined as
	\begin{equation}
	\chi_{\mathrm{geom}}=\arg\sp{\psi_{0}}{\tilde{\psi}_{N}}=\arg\bra{\psi_{0}}\mathcal{M}_{N}^{(r_{N})}\dots\mathcal{M}_{1}^{(r_{1})}\ket{\psi_{0}}.\label{eq:GP_def_WM}
	\end{equation}
	
\begin{figure}
		\begin{centering}
			\includegraphics[width=1\columnwidth]{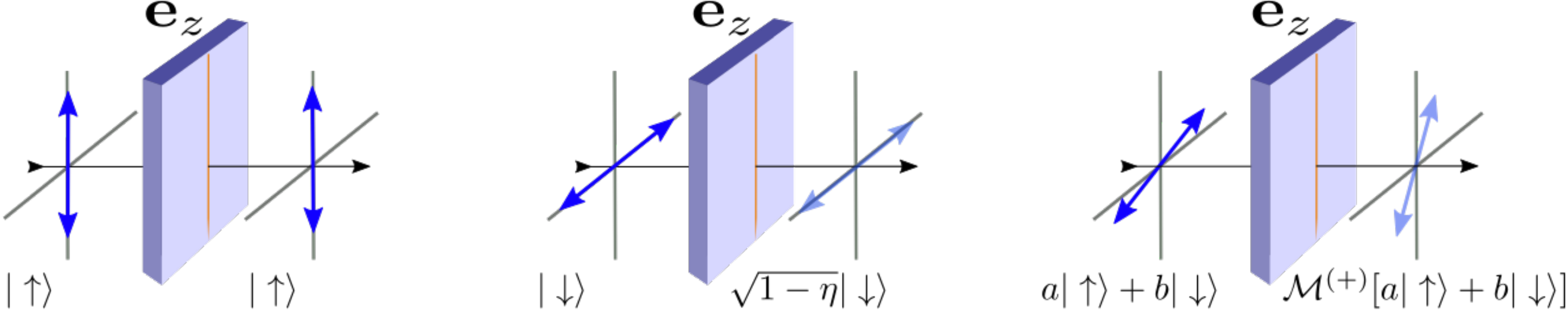}
			\par\end{centering}
		\caption{\label{fig:polarizer_as_detector}\textbf{An imperfect polarizer implementing
				a null-type weak measurement.} Each polarizer transmits a certain
			polarization (here, $\mathbf{e}_{z}$), with certainty. An impinging
			beam with generic polarization (blue arrows) is either transmitted,
			resulting in a null readout, $r=+$ , or absorbed, $r=-$. (a) A vertically
			polarized photon ($\protect\ket{\uparrow}$) is transmitted without
			altering its polarization; (b) a horizontally polarized photon ($\protect\ket{\downarrow}$)
			is transmitted with a probability $\text{1-\ensuremath{\eta}}<1$
			(pale blue arrow); (c) a photon of generic polarization $\protect\ket{\psi}=a\protect\ket{\uparrow}+b\protect\ket{\downarrow}$
			is transmitted with probability $\protect\abs a^{2}+(1-\eta)\protect\abs b^{2}$
			and a modified polarization state , $\vert\psi\rangle\to M_{\eta}(\mathbf{e}_{z},r=+)\vert\psi\rangle=\mathcal{M}^{(+)}\protect\ket{\psi}$.
			By rotating the polarizers and adding phase plates (e.g., quarter-wave
			plates), it is possible to engineer a fully transmitted polarization
			direction $\vert+\mathbf{n}\rangle=\cos\theta/2\protect\ket{\uparrow}+e^{i\varphi}\sin\theta/2\protect\ket{\downarrow}$.}
	\end{figure}

	In order to further proceed with our analysis we need to specify the
	nature of the measurement. For the present study we opt for so-called
	null-type weak measurements \cite{Zilberberg2013,Zilberberg2014,Ruskov2007}. The
	state before the measurement is $\ket{\psi}=a\ket{\mathbf{n}}+b\ket{-\mathbf{n}}$,
	where $\sigma{}_{\mathbf{n}}\ket{\pm\mathbf{n}}=\pm\ket{\pm\mathbf{n}}$,
	while the detector initial state is $\ket +$. The measurement process
	is mediated by the system-detector interaction which is described
	by the interaction-induced mapping
	\begin{multline}
	\left(a\ket{\mathbf{n}}+b\ket{-\mathbf{n}}\right)\ket +\\
	\to\left(a\ket{\mathbf{n}}+b\sqrt{1-\eta}\ket{-\mathbf{n}}\right)\ket ++b\sqrt{\eta}\ket{-\mathbf{n}}\ket -.\label{eq:state-evol}
	\end{multline}
	Following this step, the detector is projectively measured in the
	basis of $\ket{\pm}$ states. Note that this measurement protocol
	has the following properties: (i) if the initial system state is $\ket{\mathbf{n}}$
	($a=1$, $b=0$ in Eq.~(\ref{eq:state-evol})), it gives with certainty
	readout $r=+$ and does not alter the state of the system; (ii) if
	the initial system state is $\ket{-\mathbf{n}}$ ($a=0$, $b=1$ in
	Eq.~(\ref{eq:state-evol})), it yields readouts $r=-$ or $r=+$
	with probabilities $p_{-}=\eta$ and $p_{+}=1-\eta$, respectively,
	again without altering the state of the system. In general, when the
	system state is in a superposition of $\ket{\mathbf{n}}$ and $\ket{-\mathbf{n}}$,
	the measurement \emph{does} alter the system state. For $\eta\ll1$,
	the detector remains practically always in its initially prepared
	state ($r=+$, i.e., null-outcome), modifying the system state only
	slightly; yet with probability $\eta\abs b^{2}$ the readout is $r=-$,
	inducing a jump in the system state to $\ket{-\mathbf{n}}$. Considering
	only the experimental runs resulting in $r=+$ allows one to define
	``null weak values'' \cite{Zilberberg2013,Zilberberg2014}. For arbitrary
	$\eta$, such postselected measurements may be implemented, with imperfect
	polarizers, as depicted in Fig.~\ref{fig:polarizer_as_detector}:
	a photon of one polarization is always transmitted ($r=+$), while
	a photon of the orthogonal polarization has finite probability to
	be transmitted ($r=+$) or absorbed ($r=-$). Here, we do not restrict
	ourselves to postselected measurements and to the $\eta\ll1$ limit.
	Below (cf. Methods) we address a Hamiltonian implementation of such
	measurements in the spirit of the von Neumann \cite{Mello2014} measurement
	model.
	
	Define the normalized state $\ket{\psi_{k}}=\vert\tilde{\psi}_{k}\rangle/\sqrt{\sp{\tilde{\psi}_{k}}{\tilde{\psi}_{k}}}$.
	With the standard parametrization, $\vert\psi_{k}\rangle=e^{i\alpha_{k}}(\cos\Theta_{k}/2\ket{\uparrow}+e^{i\varPhi_{k}}\sin\Theta_{k}/2\ket{\downarrow})$,
	the sequence of states is mapped onto a discrete trajectory on the
	Bloch sphere with spherical coordinates $\Theta_{k}$ and $\varPhi_{k}$.
	Fig.~\ref{fig:examples_of_trajectories}(right) depicts state trajectories
	that correspond to measurement orientation sequences (Fig.~\ref{fig:examples_of_trajectories}(left))
	of various measurement strengths. The particular type of measurement
	we employ guarantees that $\sp{\tilde{\psi}_{k}}{\tilde{\psi}_{k-1}}=\bra{\tilde{\psi}_{k-1}}\mathcal{M}_{k}^{(r_{k})\dagger}\ket{\tilde{\psi}_{k-1}}=\left(\bra{\tilde{\psi}_{k-1}}\mathcal{M}_{k}^{(r_{k})}\ket{\tilde{\psi}_{k-1}}\right)^{*}>0$,
	enabling us to express the above geometric phase in the same form
	as the Pancharatnam phase (\ref{definitionphase}). It thus follows
	that $\chi_{\mathrm{geom}}=-\Omega/2$ can be expressed via the solid
	angle $\Omega$ subtended by a piecewise trajectory on the Bloch sphere
	that connects the neighboring states ($\ket{\psi_{k}}$ and $\ket{\psi_{k+1}}$;
	here we imply $\ket{\psi_{N+1}}\coloneqq\ket{\psi_{0}}$) along shortest
	geodesics. Note the difference between weak and projective measurements.
	In the latter, the system states $\ket{\psi_{k}}$ are fully determined
	by the measurement orientation and the measurement readout $r_{k}$.
	By contrast, the system state following a weak measurement also depends
	on its strength $\eta<1$ and on the state before the measurement.
	Furthermore, for a quasicontinuous sequence of strong measurements
	($N\rightarrow\infty$ and $\lVert\mathbf{n}_{k+1}-\mathbf{n}_{k}\rVert=O(1/N)$),
	the readout $r_{k}=-$ is impossible due to the dynamical quantum
	Zeno effect \cite{Facchi1999}, rendering all readouts $r_{k}=+$ and
	the measurement-induced trajectory deterministic. For a quasicontinuous
	sequence of \emph{weak} measurements, the trajectory is, instead,
	stochastic, manifested in a variety of possible readout sequences
	$\left\{ r_{k}\right\} $, cf.~Fig.~\ref{fig:examples_of_trajectories}(right).
	The probability of obtaining a specific sequence of readouts $\left\{ r_{k}\right\} $
	is given by $P_{\left\{ r_{k}\right\} }=\sp{\tilde{\psi}_{N}}{\tilde{\psi}_{N}}$.
	
	For a general choice of $\eta$ the final state $\vert\tilde{\psi}_{N}\rangle$
	may not be proportional to the initial state $\vert\psi_{0}\rangle$,
	meaning the trajectory $\ket{\psi_{1}}\rightarrow...\rightarrow\ket{\psi_{N}}$
	is not closed. For simplicity, we take the last measurement to be
	strong ($\eta_{N}=1$) and \emph{postselect} it to yield $r_{N}=+$
	(i.e., discard those experimental runs that yield $r_{N}=-$), hence
	$\mathcal{M}_{N}^{(r_{N})}=\mathcal{M}_{N}^{(+)}=\ket{\psi_{0}}\bra{\psi_{0}}=\mathcal{P}_{0}$.
	The probability of getting a specific sequence of readouts $\left\{ r_{k}\right\} $
	can then be expressed as
	\begin{equation}
	P_{\left\{ r_{k},r_{N}=+\right\} }=\abs{\langle\psi_{0}|\tilde{\psi}_{N}\rangle}^{2}=\abs{\langle\psi_{0}|\tilde{\psi}_{N-1}\rangle}^{2}\label{definitionprobability}
	\end{equation}
	with $\ket{\tilde{\psi}_{N}}=\ket{\tilde{\psi}_{\left\{ r_{k}\right\} }}$
	as defined in Eq.~(\ref{eq:psif}). Thus,
	\begin{equation}
	\bra{\psi_{0}}\mathcal{M}_{N-1}^{(r_{N-1})}\dots\mathcal{M}_{1}^{(r_{1})}\ket{\psi_{0}}=\sqrt{P_{\left\{ r_{k},r_{N}=+\right\} }}e^{i\chi_{\mathrm{geom}}}.\label{eq:phase_prob_matrix_element}
	\end{equation}
	
	We next study $\chi_{\mathrm{geom}}$ as a function of the measurement
	strength. We consider $N\rightarrow\infty$ measurements with measurement
	orientations $\left\{ \mathbf{n}_{k}\right\} $ that follow a given
	parallel on the sphere, $(\theta_{k},\varphi_{k})=(\theta,2\pi k/N)$.
	The initial state is $\ket{\psi_{0}}=\cos\theta/2\ket{\uparrow}+\sin\theta/2\ket{\downarrow}$,
	cf.~Fig.~\ref{fig:examples_of_trajectories}(right). The measurement
	strength of each individual measurement is $\eta_{k}=\eta=4c/N\to0$
	with $c$ being a non-negative constant (except for $\eta_{N}=1$).
	The sequence of $N-1$ weak measurements can be characterized by an
	effective measurement strength $\eta_{\mathrm{eff}}=1-e^{-4c}$, $0\leqslant\eta_{{\rm eff}}\leqslant1$.
	
	\section*{Probability distribution of the measurement-induced geometric phase}
	
	\begin{figure}
		\centering{}\includegraphics[width=1\columnwidth]{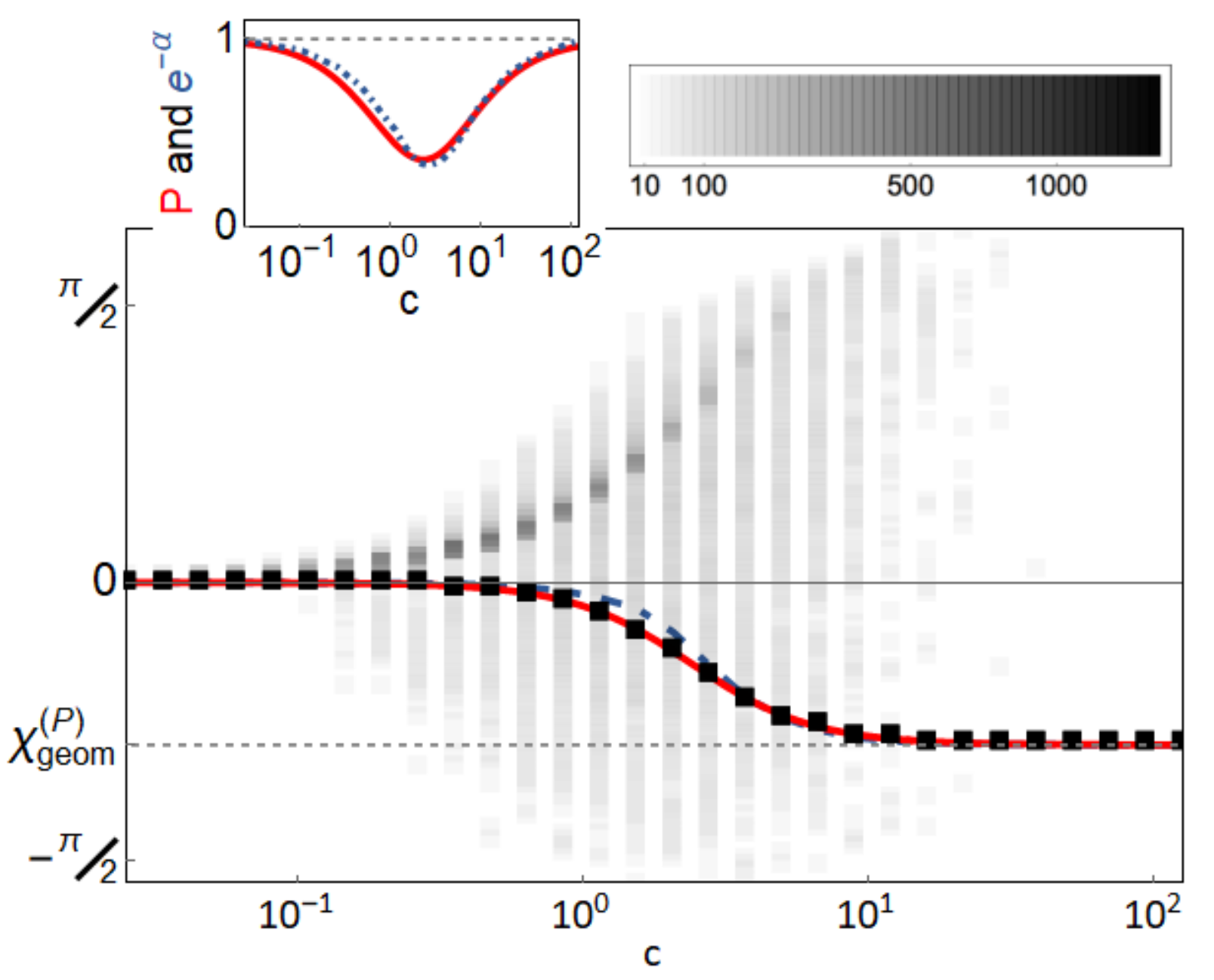} \caption{\textbf{\label{fig:geom_phase_statistics}Statistics of the measurement
				induced geometric phase. }The geometric phase induced by continuous
			measurements ($\theta=\pi/4$) as a function of the parameter $c$
			controlling the strength of measurements: For the $\{r_{k}=+\}$ postselected
			trajectory, $\chi_{\mathrm{geom}}$ is given by Eq.~(\ref{postselectedresult})
			(red solid line). The averaged geometric phase, $\bar{\chi}_{\mathrm{geom}}$
			Eq.~(\ref{defaverage}), is represented by the blue dot-dashed line.
			Both curves show a similar transition from a finite geometric phase
			for strong measurements to a vanishing one in the weak measurement
			limit. The gray-scale density plot shows the probability distribution
			of the geometric phases in the absence of postselection. Note that
			the phase of the postselected trajectory is the most probable one
			(black squares). For $c\simeq0.1\ldots1$, a secondary peak emerges,
			and persists for intermediate measurement strengths before fading
			out again. In the weak measurement regime, the $r_{k}=-$ readouts
			are not probable due to the weakness of the measurement. The secondary
			peak seems to arise due to the trajectories with a single $r_{k}=-$
			readout at $k\sim N/2$, where the probability of such a readout is
			the highest. At intermediate measurement strengths, the distribution
			is dichotomic and broader as the probability of $r_{k}=-$ outcomes
			is higher. For large $c$, the secondary peak is suppressed and the
			distribution finally collapses onto the result enforced by the dynamical
			quantum Zeno effect. Inset: The probability of observing the $\{r_{k}=+\}$
			readout sequence (red solid line) and the suppression factor $e^{-\alpha}$
			in Eq.~(\ref{defaverage}) (blue dot-dashed line) arising due to
			averaging ($\theta=\pi/4$). The probability distribution and averaging
			were obtained performing Monte Carlo simulations with $N=500$ measurements
			per sequence and $N_{\mathrm{realizations}}=4000$ realizations.}
	\end{figure}
	
	\textcolor{black}{The probability distribution of the geometric phases
		is reported in Fig.~\ref{fig:geom_phase_statistics} as a function
		of the parameter $c$ quantifying the effective measurement strength.
		For $c\to0$, the distribution is peaked around $\chi_{{\rm geom}}=0$
		corresponding to a vanishing backaction from the measurement process.
		With increasing measurement strength, the distribution develops a
		main peak which continuously evolves towards the Pachnaratnam phase
		in the strong measurement limit. This peak corresponds to the the
		most probable trajectory associated with a specific readout sequence,
		$r_{k}=+$ for all $k$, cf.~Fig.~\ref{fig:examples_of_trajectories}
		(red solid line). A secondary peak develops for intermediate measurement
		strengths due to the non-vanishing probability of obtaining $r_{k}=-$
		for some $k$.}
	
	We first turn our attention to the geometric phase associated with
	a specific readout sequence, $r_{k}=+$ for all $k$ (this means that
	if any $r_{k}=-$, that particular experimental run should be discarded).
	For a generic measurement strength, it is the most probable measurement
	outcome, hence the corresponding GP is the most likely one. We parametrize
	the Hilbert space trajectory as $\ket{\psi(t)}$, $t\in[0,2\pi]$,
	so that $\ket{\psi(t=\pi k/N)}=\ket{\psi_{k}}$ for $k=1,...,N-1$
	and $\ket{\psi(t\in[\pi,2\pi])}$ is the shortest Bloch sphere geodesic
	between $\ket{\psi_{N-1}}$ and $\ket{\psi_{N}}=\ket{\psi_{0}}$,
	cf.~Fig.~\ref{fig:examples_of_trajectories}(right). This parametrization
	results in a quasi-continuous trajectory since $\lVert\vert\psi_{k+1}\rangle-\vert\psi_{k}\rangle\rVert=\mathcal{O}(1/N)$
	for $k<N-1$. We investigate the behavior of $\chi_{\mathrm{geom}}$
	and link it to the behavior of $\ket{\psi(t)}$ as a function of the
	measurement strength $\eta$. Since the measurements are not projective
	(measurement strength $\eta\to0$), the state after each measurement
	is not necessarily the $\uparrow$-eigenstate of $\sigma_{\mathbf{n}_{k}}$.
	The state trajectory on the Bloch sphere for $\theta=\pi/4$ and $c=3$
	is shown in Fig.~\ref{fig:examples_of_trajectories} (red solid line).
	The probability $P=P_{\left\{ r_{k}=+\right\} }$ of measuring the
	desired readouts and the corresponding geometric phase $\chi_{\mathrm{geom}}$
	(cf.~Eq.~(\ref{eq:phase_prob_matrix_element})) can be calculated
	analytically for $N\to\infty$ and are given by
	\begin{equation}
	\sqrt{P}e^{i\chi_{\mathrm{geom}}}=-e^{-c}(\cosh(\tau)+z\sinh(\tau)/\tau),\label{postselectedresult}
	\end{equation}
	with $\tau=\sqrt{z^{2}-\pi^{2}\sin^{2}\theta}$ and $z=c+i\pi\cos\theta$,
	cf.~Fig.~\ref{fig:geom_phase_statistics} (red solid lines).
	
	We note three qualitatively different regimes depending on the parameter
	$c$ controlling the effective measurement strength. For strong measurements
	($c\rightarrow\infty$), one obtains Zeno-like dynamics: the state
	follows meticulously the measurement orientation. In this limit, the
	probability of the successful postselection of the measurement readouts
	approaches $1$ and the GP is $-\Omega/2$: half the solid angle enclosed
	by the measurement orientation, $\Omega=2\pi(1-\cos\theta)$. Similarly,
	in the infinitely weak measurement limit ($c\rightarrow0$), the probability
	of obtaining all readouts $r_{k}=+$ approaches $1$. In this limit,
	however, the result stems from the fact that the system barely interacts
	with the detector (cf. Eq. (\ref{eq:state-evol})); $r_{k}=+$ is
	the only possible measurement readout and back-action is practically
	absent: the system remains in its initial state at all times and accumulates
	no geometric phase. Finally, for intermediate strength measurements,
	the system reacts to the measurement, yet its state does not follow
	the measurement orientation but has a readout-sequence-dependent non-trivial
	trajectory. As a consequence, the trajectory with all readouts $r_{k}=+$
	occurs with reduced probability and a smaller postselected geometric
	phase as compared with the strong measurement limit.
	
	As a follow up, we characterize the effect of the probability distribution
	of $\chi_{{\rm geom}}$ by studying the average GP and its behavior
	as a function of $\eta.$ We define the averaged geometric phase $\bar{\chi}_{{\rm geom}}$
	as
	\begin{align}
	e^{2i\bar{\chi}_{\mathrm{geom}}-\alpha}:=&\langle e^{2i\chi_{\mathrm{geom}}}\rangle_{\mathrm{realizations}}\notag \\
	=&\sum_{\{r_{k}\}}\left(\bra{\psi_{0}}\mathcal{M}_{N-1}^{(r_{N-1})}\dots\mathcal{M}_{1}^{(r_{1})}\ket{\psi_{0}}\right)^{2},\label{defaverage}
	\end{align}
	\textcolor{black}{motivated by physically measurable observables (cf.
		Supplementary Material).} Here, the sum extends over all possible
	measurement readouts $\{r_{k}\}$. Also for the averaged phase, one
	distinguishes three qualitatively different measurement regimes (cf.
	Fig. \textcolor{blue}{\ref{fig:geom_phase_statistics}}). In the limits
	of either strong or weak measurements, all trajectories except the
	one corresponding to $r_{k}=+$ for all $k$ carry negligible probabilities.
	Therefore, the average GP approaches the value computed for the postselected
	trajectory in those limits. The absence of fluctuations in the GP
	and the near certainty of successful postselection in the final measurement
	implies that dephasing is absent ($\alpha\rightarrow0$). Only in
	the intermediate regime, the distribution of trajectories is broadened;
	the average GP then noticeably deviates from the one in the postselected
	trajectory and dephasing emerges. Interestingly, the dephasing suppression
	factor accompanying the geometric phase follows the same behavior
	as the probability suppression of the postselected trajectory considered
	above. This indicates the fact that the postselected trajectory yields
	the dominant contribution, being often the most probable trajectory.
	Indeed, while the probability of this trajectory is $\mathcal{O}(1)$,
	all other trajectories contribute each with its own phase and a small
	weight.

	\section*{Postselected geometric phase: topological nature of strong to weak
		measurement transition}
	
	\begin{figure}
		\begin{centering}
			\includegraphics[width=1\columnwidth]{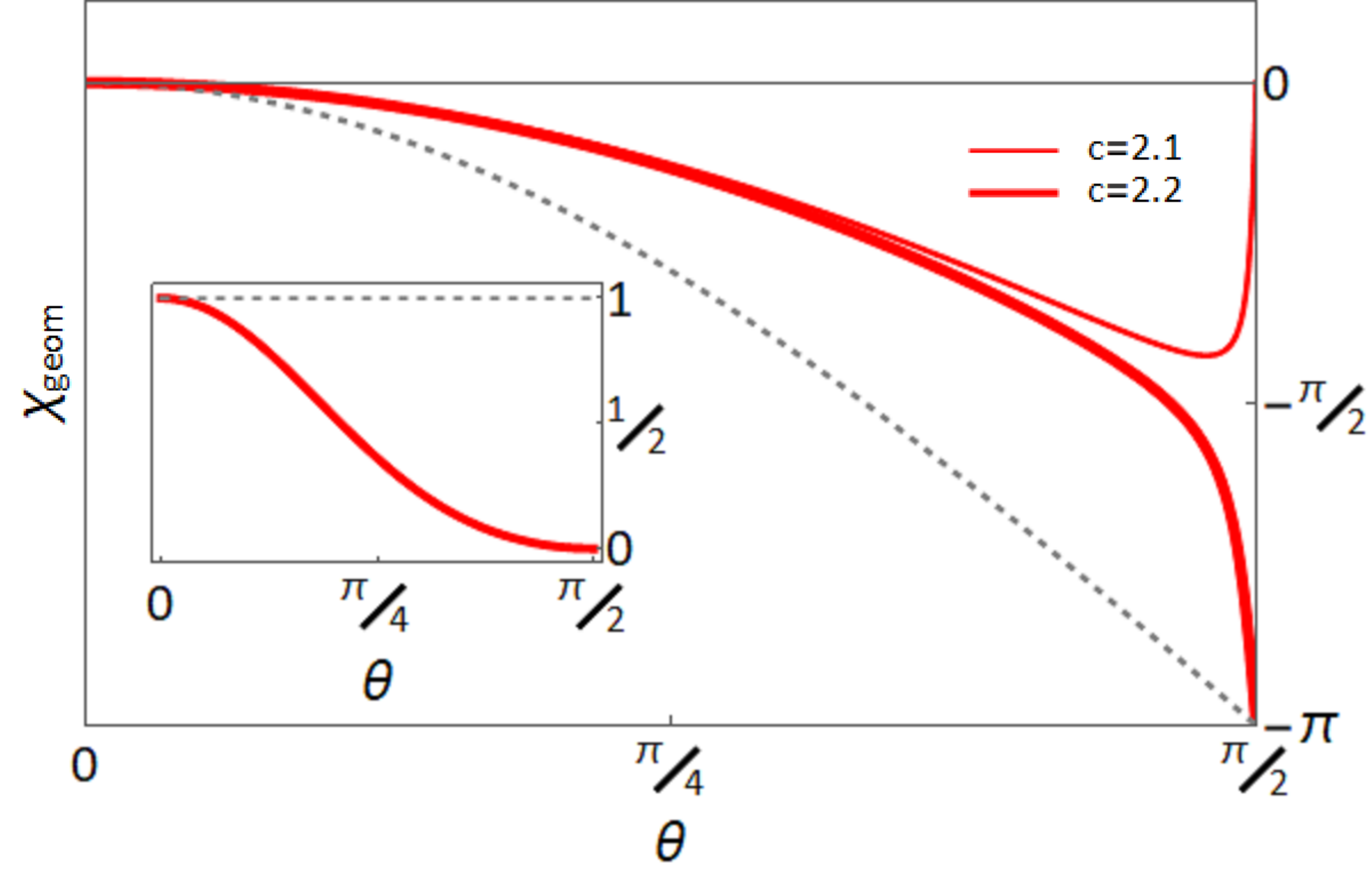}
			\par\end{centering}
		\caption{\textbf{\label{fig:theta_dependence}Non-monotonicity of geometric
				phases}. Dependence of the geometric phase on the polar angle $\theta$
			for the postselected trajectory (red solid lines) for different values
			of the integrated measurement strength (cf. legend). The ideal strong
			measurement dependence for $c\rightarrow\infty$ is presented as a
			grey dashed line. The asymptotic dependence of the GP on $\theta$
			displays an abrupt transition from monotonic to non-monotonic behavior
			in the vicinity of $c=2.15$. The behavior is underlined by the fact
			that $\chi_{{\rm geom}}(\pi/2)$ can assume only discrete values,
			$0$ or $-\pi$. Inset: the probability of observing the most probable
			trajectory with postselected readout sequence $\{r_{k}=+\}$ at $c=2.1$;
			the grey dashed line indicates $P=1$ for $c\rightarrow\infty$, showing
			the dynamical quantum Zeno effect.}
	\end{figure}
	
	We next study the dependence of the GP of the postselected trajectory
	with all outcomes $r_{k}=+$ on the measurement sequence polar angle
	$\theta$ for a given measurement strength. Consider the continuous
	function $\chi_{{\rm geom}}(\theta):\:[0,\pi]\to\mathbb{R}$. (Note
	that although $\chi_{{\rm geom}}$ is a phase and is thus defined
	$\mod2\pi$, we unfold it to have values in $\mathbb{R}$ by demanding
	that $\chi_{{\rm geom}}(0)=0$ and that $\chi_{{\rm geom}}(\theta)$
	is continuous). For $c\gg1$ (i.e., $\eta_{{\rm eff}}\to1$), it behaves
	as the standard Pancharatnam-Berry phase, $\chi_{{\rm geom}}(\theta)=\pi(\cos\theta-1)$.
	For $c=0$, $\chi_{{\rm geom}}(\theta)=0$. We find that the regimes
	of infinitely weak and infinitely strong measurements are separated
	by a sharp transition, cf.~Fig.~\ref{fig:theta_dependence}. For
	$\theta=\pi/2$, the expression in the r.h.s. of Eq.~(\ref{postselectedresult})
	is real, implying that $\chi_{{\rm geom}}(\theta=\pi/2)$ can only
	take values $0$ and $-\pi$. Thus, the interpolation of $\chi_{{\rm geom}}(\pi/2)$
	between the infinitely weak and infinitely strong regimes must involve
	a discontinuous jump. As the measurement strength is reduced, we have
	$\chi_{{\rm geom}}(\pi/2)=-\pi$ to a critical strength, $c_{\mathrm{crit}}\approx2.15$,
	below which $\chi_{{\rm geom}}(\pi/2)=0$. Note also that $\chi_{{\rm geom}}(\theta)$
	is a monotonic function for $c>c_{\mathrm{crit}}$ and is non-monotonic
	for $c<c_{\mathrm{crit}}$. At $c=c_{\mathrm{crit}}$, $\chi_{{\rm geom}}(\pi/2)$
	is ill-defined as the probability of a successful postselection (i.e.,
	that all readouts are $r_{k}=+$) $P(\theta=\pi/2)=0$.
	
	\begin{figure}
		\centering{}\includegraphics[width=0.45\columnwidth]{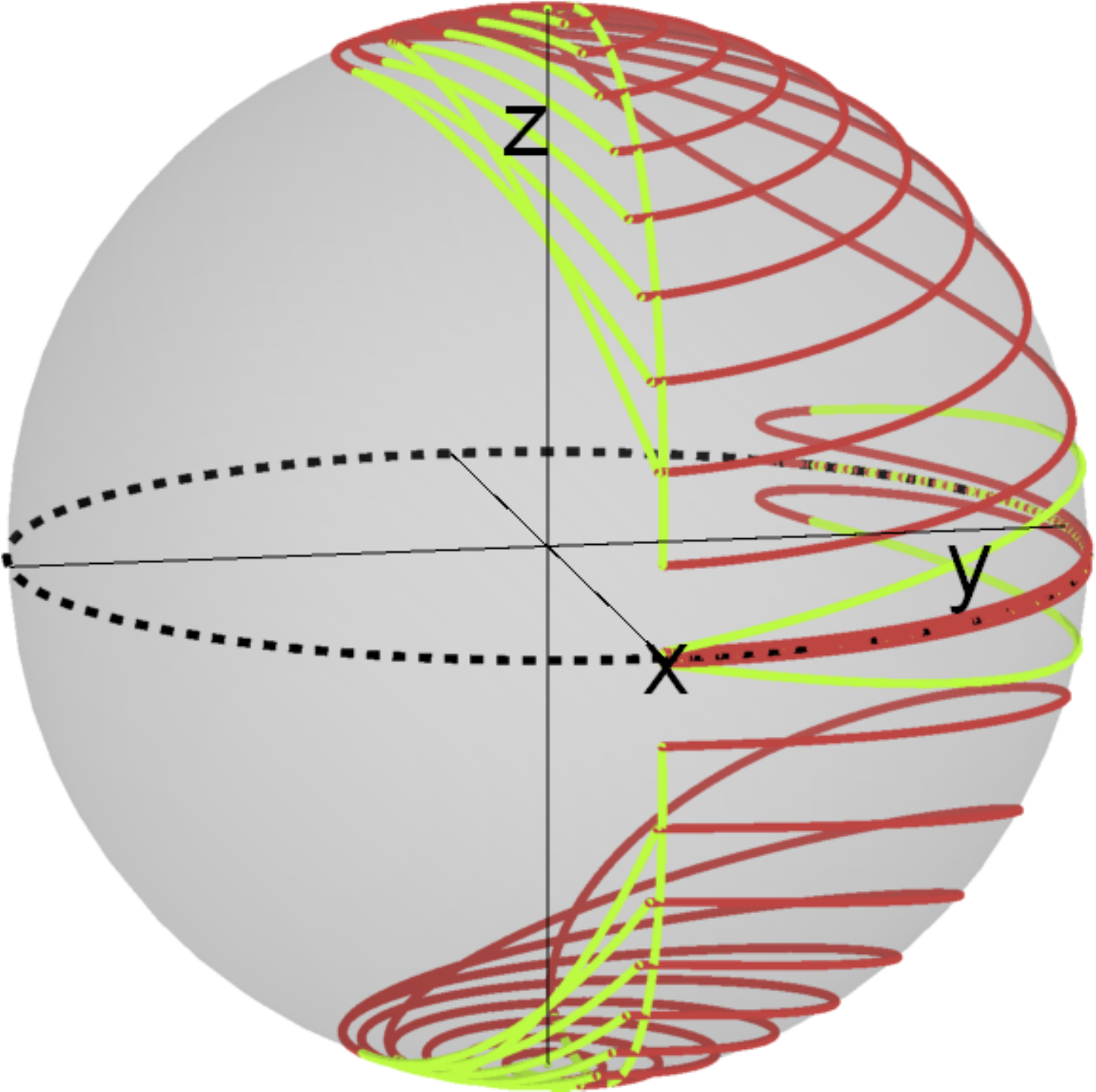}
		\includegraphics[width=0.45\columnwidth]{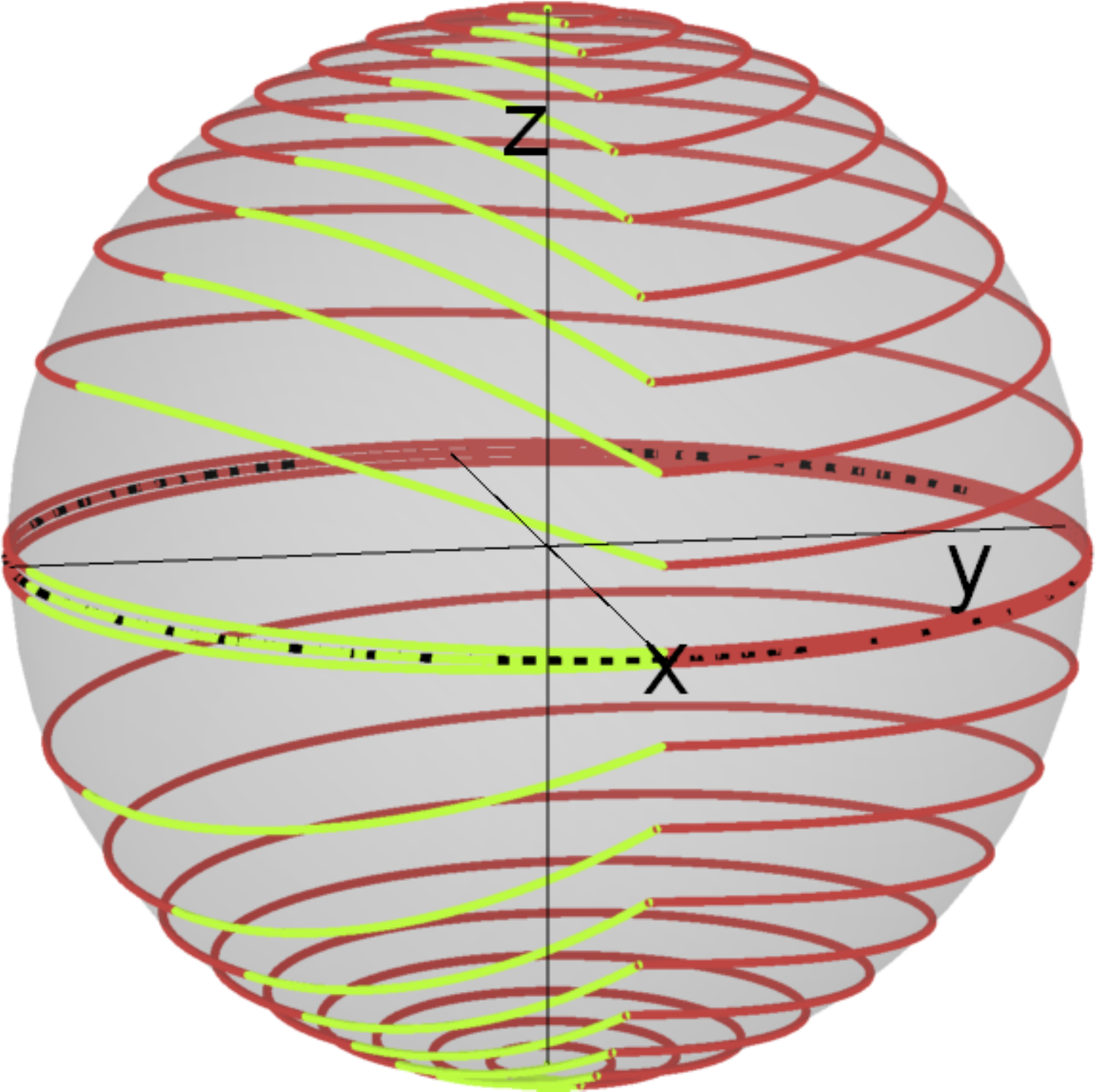} \caption{\label{fig:trajectories}\textbf{Mapping of measurement sequences
				onto the system's quantum trajectories: a topological transition.
			}A $\theta$-dependent family of trajectories on the Bloch sphere
			for $c<c_{{\rm crit}}$ (left) and $c>c_{{\rm crit}}$ (right). Yellow
			segments represent final projective measurements, ascertaining closed
			trajectories. For stronger measurements (right), the $S^{2}$ space
			of the measurement orientations ${\bf n}$ is mapped through the measurement
			process onto the whole $S^{2}$ Bloch sphere of quantum trajectories.
			For weaker measurement strengths, the $S^{2}$ space of measurement
			orientations is mapped onto a subset of the Bloch sphere. The corresponding
			Chern numbers are -1 and 0, respectively.}
	\end{figure}
	
	We relate these observations concerning the GP to the behavior of
	the induced quantum state trajectory at $\theta=\pi/2$, cf.~Fig.~\ref{fig:trajectories}.
	The quantum state trajectory for $\theta=\pi/2$ lies entirely on
	the equator of the Bloch sphere. For $c>c_{\mathrm{crit}}$, the trajectory
	after $N-1$ finite strength measurements ($\ket{\psi_{\theta=\pi/2}(t\in[0,\pi])}$)
	traverses more than half the equator; the last projective measurement
	brings it back to the original point by the shortest geodesic, completing
	the circle around the equator. The solid angle subtended by the trajectory
	is then $\Omega=2\pi$, and $\chi_{{\rm geom}}(\theta=\pi/2)=-\pi$.
	For $c<c_{\mathrm{crit}}$, $\ket{\psi_{\pi/2}(\pi)}$ has not reached
	the equator 's middle, and the last projective measurement again brings
	the system state back to the original point by the shortest geodesic,
	which in this case implies retracing its path back. The trajectory
	then subtends no solid angle, and the resulting phase $\chi_{{\rm geom}}(\theta=\pi/2)=0$.
	Note that the existence of a sharp transition at $\theta=\pi/2$ is
	protected by the fact that $M_{\eta}(\mathbf{e}_{z},r)$ is real,
	cf.~Eq.~(\ref{eq:M_down}), which guarantees that the trajectory
	always remains on the equator and thus $\chi_{{\rm geom}}(\pi/2)\in\{0,-\pi\}$.
	
	This picture, in fact, extends beyond the trajectories on the equator.
	Consider the manifold formed by \emph{all} state trajectories, which
	is obtained via measurement sequences at $\theta\in[0,\pi]$, cf.~Fig.~\ref{fig:trajectories}.
	For $c>c_{\mathrm{crit}}$, this manifold covers the Bloch sphere,
	while for subcritical $c$ it does not. The GP transition then corresponds
	to a change in the topology of the set of state trajectories ---
	thence the designation ``topological transition''. While this transition
	can be intuitively understood from the behavior of the trajectory
	on the equator, we prove it formally below (cf.~Methods) by considering
	the Chern number
	
	\begin{align}
	\mathcal{C}&\equiv\frac{1}{2\pi}\int_{0}^{\pi}d\theta\int_{0}^{2\pi}dt\thinspace\tilde{B}(\theta,t)\notag \\
	&=\frac{1}{2\pi}\left(\chi_{{\rm geom}}(\pi)-\chi_{{\rm geom}}(0)\right)\in\{0,-1\},\label{eq:chern-1}
	\end{align}
	where $\tilde{B}(\theta,t)$ is the Berry curvature
	\begin{equation}
	\tilde{B}(\theta,t)={\rm Im}\left(\partial_{t}\bra{\psi_{\theta}(t)}\partial_{\theta}\ket{\psi_{\theta}(t)}-\partial_{\theta}\bra{\psi_{\theta}(t)}\partial_{t}\ket{\psi_{\theta}(t)}\right).\label{eq:Berry_curvature}
	\end{equation}
	
    Transitions in quantum dynamics as a function of the measurement strength have been known for single qubits~\cite{Ruskov2007} and more recently discovered for many-body systems~\cite{Szyniszewski2019}. Notably, the topological nature of the transition we report here is novel. Importantly, it implies that the transition is robust against perturbing the protocol. For example, if one considers sequences of measurements that follow generic closed curves different from the parallels considered above, it would not be possible to define $\chi_{{\rm geom}}(\pi/2)$ and determine the transition from its discrete set of values. However, as long as the family of measurement sequences wraps the sphere, there is a Chern number which assumes a discrete set of values (characterizing a global property of the set of measurement-induced trajectories) controlled by the measurement strength. Importantly, the Chern number is different in the limits of weak and strong measurements. This guarantees the existence of a critical measurement strength, $c_\mathrm{crit}$, at which the Chern number changes abruptly, and a concurrent jump of the phase $\chi_\mathrm{geom}$ associated with a critical measurement sequence.  Unlike the transition, whose existence is protected by the change of the topological invariant, the value $c_{\mathrm{crit}}$ at which it takes place and the corresponding critical measurement sequence are non-universal and depend on the specifics of the protocol.

	\section*{Experimental implementations}
	
		\begin{SCfigure*}[\sidecaptionrelwidth][t]
		\centering
		\includegraphics[width=11.4cm]{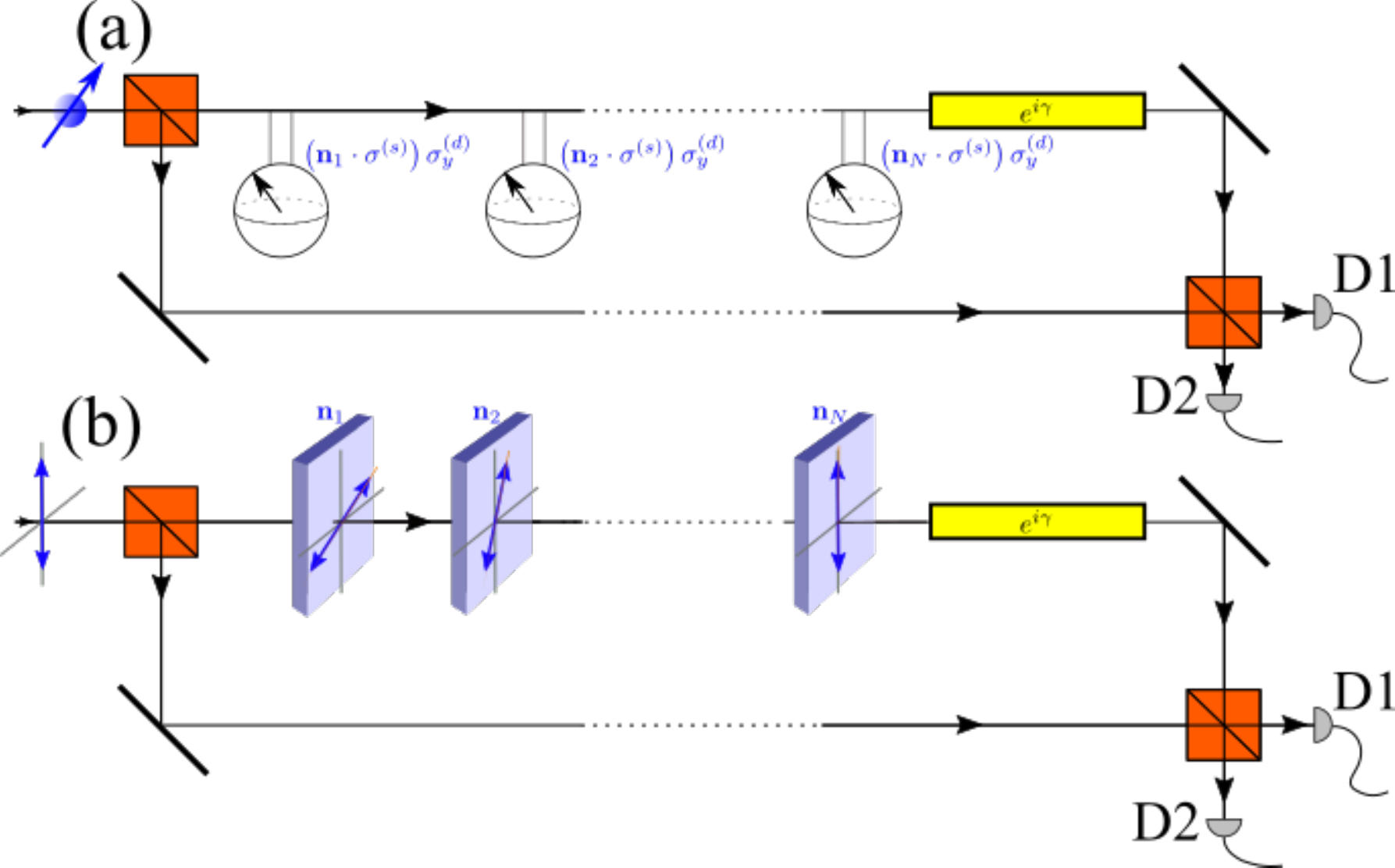}
		\caption{\textbf{Experimental setups for observing
				measurement-induced GPs}. (a) Observing the postselected GP in a Mach-Zehnder
			interference setup. A particle interacts with a sequence of detectors
			in one arm. The null-type of the measurements employed means that
			the detectors do not change their state for $r_{k}=+$ readouts. No
			\textquotedbl which-path\textquotedbl{} information is implied, hence
			the GP acquired in the postselected readout sequence $\left\{ r_{k}=+\right\} $
			is manifest in the interference pattern at drains D1 and D2. (b) An
			equivalent setup with polarized photons as particles and imperfect
			polarizers acting as postselected weak measurements. In both setups,
			we assume an extra phase difference $e^{i\gamma}$ produced by means
			other than measurements.}
		\label{fig:interferometers}
	\end{SCfigure*}

	In order to detect the postselected GP in an experiment, we design
	a protocol based on a Mach-Zehnder interferometer incorporating detectors
	in one of its arms (cf.~Fig.~\ref{fig:interferometers}(a)). An
	impinging particle with an internal degree of freedom (spin for electrons,
	polarization for photons) in state $\ket{\psi_{0}}$ is split into
	two modes in the two interferometer arms. The compound system-detector
	state is then $\ket{\Psi_{i}}=\vert\psi_{0}\rangle\otimes(\ket{a=1}+\ket{a=-1})\otimes\ket{+...+}/\sqrt{2}$,
	where $a=\pm1$ describes the particle being in the upper or lower
	arm, respectively, and $\ket{+...+}$ is the initial state of the
	detectors. In the upper arm, the particle is subsequently measured
	by all the detectors; in addition it acquires an extra dynamical phase
	$\gamma$ controlling the interference. Running through the lower
	arm, the state is left untouched. Traversing the interferometer, the
	state is then $\ket{\Psi_{f}}=\vert\psi_{0}\rangle\ket{a=-1}\ket{+...+}/\sqrt{2}+e^{i\gamma}\sum_{\left\{ r_{k}\right\} }\prod_{k=1}^{N}\mathcal{M}_{k}^{(r_{k})}\vert\psi_{0}\rangle\ket{a=1}\ket{\left\{ r_{k}\right\} }/\sqrt{2}$,
	where $r_{k}$ is the readout of the $k$-th measurement, and $\ket{\left\{ r_{k}\right\} }$
	is the corresponding collective state of all the detectors. The state
	with all readouts $r_{k}=+$ coincides with the initial state of all
	the detectors $\ket{+...+}$, therefore producing interference \footnote{Note that in this scheme we actually do not postselect the $\{r_{k}=+\}$
		readout sequence but since only this sequence yields no \textquotedbl which-path\textquotedbl{}
		information, it is the only one giving rise to interference. Therefore,
		in the final interference, we only see the geometric phase of the
		$\{r_{k}=+\}$ trajectory.}. The intensities observed at drains D1 and D2 are
	\begin{align}
	I_{1,2}&=\frac{I_{0}}{2}\left(1\pm\mathrm{Re}\thinspace e^{i\gamma}\bra{\psi_{0}}\prod_{k=1}^{N}\mathcal{M}_{k}^{(+)}\vert\psi_{0}\rangle\right)\notag\\
	&=\frac{I_{0}}{2}\left(1\pm\sqrt{P}\mathrm{Re}\thinspace e^{i\chi_{\mathrm{geom}}+i\gamma}\right),\label{eq:postselected_intensity_1}
	\end{align}
	where, for $N\rightarrow\infty$, $\sqrt{P}e^{i\chi_{\mathrm{geom}}}$
	is given by Eq.~(\ref{postselectedresult}) and $I_{0}$ is the intensity
	of the incoming particle beam. The probability of a successful postselection
	$P=P_{\left\{ r_{k}=+\right\} }$ thus determines the interference
	visibility, and the weak-measurement-induced phase $\chi_{\mathrm{geom}}$
	is directly related to the interference phase. Note that the null-type
	measurements are essential here as they induce backaction without
	forming a \textquotedbl which-path\textquotedbl{} signature, thus facilitating
	interference.
	
	In practice, this protocol can be implemented employing imperfect
	optical polarizers (cf.~Fig.~\ref{fig:interferometers}(b)). An
	imperfect polarizer can transmit (readout $+$) or absorb (readout
	$-$) the impinging light. More specifically, each polarizer fully
	transmits one given polarization ($\vert+\mathbf{n}_{k}\rangle$)
	and partially absorbs the orthogonal one ($\vert-\mathbf{n}_{k}\rangle$).
	A transmitted beam corresponds to a $+$ readout, thus implementing
	a postselected null-type measurement considered above (cf. Eq. (\ref{eq:state-evol})).
	By rotating the polarizers and adding phase plates (e.g., quarter-wave
	plates), it is possible to control the orientation $\vert+\mathbf{n}_{k}\rangle$
	that is fully transmitted. The larger the probability to absorb a
	photon of polarization $\vert-\boldsymbol{n}_{k}\rangle$, the stronger
	the measurement is. The polarization state of a beam traversing the
	sequence of polarizers reproduces the postselected state with all
	readouts $r_{k}=+$. Installing a set of polarizers in one of the
	interferometer's arms would allow us to detect the postselected GP
	through the interference pattern. The obtained signals at the interferometer
	outputs D1 and D2 are
	\begin{align}
	I_{1,2}&=I_{0}\left(\frac{1+P}{4}\pm\frac{1}{2}\sqrt{P}\mathrm{Re}\thinspace e^{i\chi_{\mathrm{geom}}+i\gamma}\right)\notag \\
	&=\frac{I_{0}}{4}\abs{1\pm\sqrt{P}e^{i\chi_{\mathrm{geom}}+i\gamma}}^{2}.\label{eq:postselected_intensity_2}
	\end{align}
	The GP can be extracted from the interference pattern controlled by
	$\gamma$. The difference in the intensities compared to Eq.~(\ref{eq:postselected_intensity_1})
	accounts for the loss due to the light absorption by the imperfect
	polarizers. Note that the limit of polarizers corresponding to strong
	measurements (i.e., one polarization of the beam is fully transmitted
	while the orthogonal polarization is fully blocked) is a realization
	of the Pancharatnam phase (\cite{Pancharatnam1956,Berry1996}). The polarizers
	must be carefully designed such that no additional phase difference
	between the two polarizations is accumulated by the light passing
	through a polarizer. This is particularly important because the topological
	nature of the transition investigated above is protected by the hermiticity
	of the Kraus operators, $\mathcal{M}_{k}^{(r_{k})}$. \footnote{An investigation of the weak-measurement-induced geometric phase when
		the Kraus operators are non-Hermitian will be performed elsewhere.}

    The above protocol can also be implemented using  superconducting qubit hardware \cite{Murch2013a,Weber2014,Naghiloo2019,Minev2019}. In such an  implementation, the particle's internal degree of freedom is replaced by the two lowest levels of a superconducting Josephson junction (which form the qubit). The lower reference arm of the interferometer should be then replaced by an extra level that is unaffected by the measurements.
	
	\section*{Discussion}
	
	We have shown how sequences of generalized quantum measurements may
	modify the phase of the state of the system measured, inducing a purely
	geometric phase. In other words, the trajectory traced by the quantum
	state can be directly mapped onto the phase accrued during the sequence
	of measurements. As opposed to geometric phases induced by an adiabatic
	Hamiltonian evolution, the phases obtained here depend on the measurement
	strength and are inherently stochastic. We have put forward schematic
	experimental protocols for measuring the geometric phase associated
	with a specific postselected readout sequence (in other words: of
	a specific postselected trajectory). We have shown that the mapping
	of measurement sequences to geometric phases undergoes a topological
	transition as the measurement strength is varied. This transition
	is classified through a jump of a Chern number. This transition is
	also manifest through an abrupt change of the dependence of the geometric
	phase on the polar angle, $\theta$, of the measurement sequence from
	a monotonous to a non-monotonous one. An analysis of the averaged
	geometric phases in the Supplementary material shows a similar feature
	in the $\theta$-dependence. Our analysis underscores for the first
	time the topological nature of a strong-to-weak measurement transition.

	The transition prevails in a much broader context, which is guaranteed
		by its topological nature. Since the Chern numbers in the limit of
		strong measurements and in the limit of infinitely weak measurements
		are different, the transition will take place also for measurements
		of different types, characterized by Kraus operators other than the
		ones we used (yet still Hermitian). Further, while we investigated
		quasicontinuous ($N\rightarrow\infty$) sequences of weak measurements,
		the transition will take place for any number $N\geq3$ of measurements
		(albeit the critical measurement strength will depend on $N$).
	
	We believe that the interplay between
	the topological nature of the measurement and possible topological
	structure of the system measured (associated with, e.g., non-Abelian
	quasi-particles, band structure and dynamical evolution) opens an
	intriguing horizon.

	\matmethods{
		
		\subsection*{Measurement model}
		
		The measurement sequence leading to the geometric phase in Eq. (\ref{eq:GP_def_WM})
		consists of positive-operator valued measurements (POVMs) defined
		by the Kraus operators $\mathcal{M}_{k}^{(r_{k})}=M_{\eta_{k}}(\mathbf{n}_{k},r_{k})$,
		$\vert\psi\rangle\to\mathcal{M}_{k}^{(r_{k})}\vert\psi\rangle$, as
		described in the main text. Such POVMs can be implemented with a detection
		apparatus consisting of a second qubit, whose Hilbert space is spanned
		by $\left|+\right\rangle $ and $\left|-\right\rangle $ and which
		is coupled to the system via the Hamiltonian
		\begin{equation}
		H_{{\bf n}}(t)=\lambda(t)(1-\sigma_{\mathbf{n}}^{(s)})\sigma_{y}^{(d)}/2.\label{eq:H}
		\end{equation}
		Here, $\sigma^{(s/d)}$ denote the Pauli matrices acting on the system
		and detector, respectively, $\sigma_{\mathbf{n}}=\mathbf{n}\cdot\boldsymbol{\sigma}$
		and $\mathbf{n}=(\sin\theta\cos\varphi,\sin\theta\sin\varphi,\cos\theta)$,
		$0\leqslant\theta\leqslant\pi$, $0\leqslant\varphi<2\pi$, defines
		the measurement direction. The system and detector are initially ($t=0$)
		decoupled in the state $|\psi_{s}^{(\mathrm{in})}\rangle\otimes\left|+\right\rangle $,
		where
		\begin{equation}
		|\psi_{s}^{(\mathrm{in})}\rangle=a\ket{\uparrow}+b\ket{\downarrow}=\begin{pmatrix}a\\
		b
		\end{pmatrix}.\label{eq:spin_states_as_column_vectors}
		\end{equation}
		The measurement coupling $\lambda(t)\neq0$ is then switched on for
		$t\in\left[0,T\right]$ to obtain the entangled state:
		\begin{align}
		|\psi_{\mathrm{ent}}\rangle&=\exp\left[-ig(1-\sigma_{\mathbf{n}}^{(s)})\sigma_{y}^{(d)}/2\right]|\psi_{s}^{(\mathrm{in})}\rangle\left|+\right\rangle \notag\\
		&=M_{\eta}(\mathbf{n},+)|\psi_{s}^{(\mathrm{in})}\rangle\left|+\right\rangle +M_{\eta}(\mathbf{n},-)|\psi_{s}^{(\mathrm{in})}\rangle\left|-\right\rangle .
		\end{align}
		Here, $g=\int_{0}^{T}dt\lambda(t)$ determines the measurement strength
		$\eta=\sin^{2}g$. The matrices $M_{\eta}(\mathbf{n},+)$ and $M_{\eta}(\mathbf{n},-)$
		are defined by
		\begin{equation}
		M_{\eta}(\mathbf{n},r)=R^{-1}(\mathbf{n})M_{\eta}(\mathbf{e}_{z},r)R(\mathbf{n}),\label{eq:M_rotation}
		\end{equation}
		where
		\begin{equation}
		M_{\eta}(\mathbf{e}_{z},+)=\left(\begin{matrix}1 & 0\\
		0 & \sqrt{1-\eta}
		\end{matrix}\right),\,\,M_{\eta}(\mathbf{e}_{z},-)=\left(\begin{matrix}0 & 0\\
		0 & \sqrt{\eta}
		\end{matrix}\right)\label{eq:M_down}
		\end{equation}
		are the Kraus operators for the measurement orientation along the
		$z$ axis ($\mathbf{n}=\mathbf{e}_{z}$) and
		\begin{equation}
		R(\mathbf{n})=\left(\begin{matrix}\cos\theta/2 & e^{-i\varphi}\sin\theta/2\\
		\sin\theta/2 & -e^{-i\varphi}\cos\theta/2
		\end{matrix}\right)
		\end{equation}
		is a unitary matrix corresponding to the rotation of the measurement
		orientation $\vert\pm\mathbf{n}\rangle=R^{-1}(\mathbf{n})\vert\pm\mathbf{e}_{z}\rangle=R^{-1}(\mathbf{n})\vert\uparrow/\downarrow\rangle$.
		This implements a null-type weak measurement as defined in the main
		text.

			\subsection*{Geometric phase from a quasicontinuous measurement sequence and postselection}
			
			The geometric phase $\chi_{\mathrm{geom}}$ obtained from the quasicontinuous
			trajectory with all outcomes $r_{k}=+$ is given in Eq. (\ref{postselectedresult}).
			This result is obtained starting from Eq. (\ref{eq:GP_def_WM}). By
			setting $\ket{\psi_{0}}=R^{-1}(\mathbf{n}_{0})\ket{\uparrow}$, the
			readouts $r_{k}=+$ and the measurement orientations $(\theta_{k},\varphi_{k})=(\theta,2\pi k/N)$,
			and using the explicit form of Kraus operators in Eq.~(\ref{eq:M_rotation}),
			one rewrites
			
			\begin{equation}
			\bra{\psi_{0}}\mathcal{M}_{N-1}^{(+)}\dots\mathcal{M}_{1}^{(+)}\ket{\psi_{0}}=\bra{\uparrow}\delta R\left(M_{\eta=4c/N}(\mathbf{e}_{z},+)\delta R\right)^{N-1}\ket{\uparrow},\label{eq:postselected_analytics_derivation}
			\end{equation}
			where
			\begin{align}
			\delta R &= R(\mathbf{n}_{k+1})R^{-1}(\mathbf{n}_{k}) \notag \\ &= \begin{pmatrix}\cos^{2}\frac{\theta}{2}+e^{-2\pi i/N}\sin^{2}\frac{\theta}{2} & \frac{1}{2}(1-e^{-2\pi i/N})\sin\theta\\
			\frac{1}{2}(1-e^{-2\pi i/N})\sin\theta & \sin^{2}\frac{\theta}{2}+e^{-2\pi i/N}\cos^{2}\frac{\theta}{2}
			\end{pmatrix}\label{eq:dR}
			\end{align}
			is a matrix independent of $k$. The quasicontinuous limit is obtained
			by diagonalizing the $2\times2$ matrix $M_{\eta=4c/N}(\mathbf{e}_{z},+)\delta R$
			and calculating the matrix elements in Eq. (\ref{eq:postselected_analytics_derivation})
			in the limit $N\rightarrow\infty$ . This yields Eq.~(\ref{postselectedresult}).
			
			\subsection*{Chern number for the mapping of measurement orientations onto state
				trajectories}
			
			The mapping of quasicontinuous measurements orientations onto state
			trajectories is topologically classified by the Chern number in Eq.
			(\ref{eq:chern-1}). The discrete values of the Chern number are in
			correspondence with the different regimes of the $\theta$-dependence
			of $\chi_{{\rm geom}}(\theta)$. To prove this, we parametrize each
			state $\ket{\psi_{\theta}(t)}$, $\theta\in[0,\pi]$, $t\in[0,2\pi]$,
			as $\ket{\psi_{\theta}(t)}=e^{i\alpha(\theta,t)}(\cos\frac{\Theta(\theta,t)}{2}\ket{\uparrow}+\sin\frac{\Theta(\theta,t)}{2}e^{i\Phi(\theta,t)}\ket{\downarrow})=e^{i\alpha(\theta,t)}\ket{\Psi(\Theta,\Phi)}$
			with $(\Theta,\Phi)$ being coordinates on the Bloch sphere. Since
			$\ket{\psi_{\theta}(2\pi)}=\ket{\psi_{\theta}(0)}$ and $\ket{\psi_{\theta=0,\pi}(t)}=\ket{\psi_{\theta=0,\pi}(0)}$,
			the parameters $t$ and $\theta$ can be regarded as a parametrization
			of a sphere, and the map $(\theta,t)\mapsto(\Theta,\Phi)$ is equivalent
			to the mapping of a sphere to a sphere $\mathcal{F}:S^{2}\ni(\theta,t)\mapsto(\Theta,\Phi)\in S^{2}$.
			
			In the quasicontinuous limit, the Pancharatnam phase in Eq.~(\ref{definitionphase})
			reduces to the Berry phase accumulated by $\ket{\Psi}$ and can be
			computed by standard methods \cite{Berry1984a} to express it as an
			integral of the Berry curvature. In fact, for any $\theta_{0}\in[0,\pi]$,
			we have $ \chi_{\mathrm{geom}}(\theta_{0}) = \int_{0}^{\theta_{0}}\int_{0}^{2\pi}d\theta dt\thinspace\tilde{B}(\theta,t) $ (see Eq. (\ref{Eq21})).
			\begin{figure*}[bt!]
			\begin{align}
\chi_{\mathrm{geom}}(\theta_{0})=\chi_{\mathrm{geom}}(\theta_{0})-\chi_{\mathrm{geom}}(0)&=\arg\prod_{t=0}^{2\pi-dt}\sp{\Psi(\Theta(\theta_{0},t+dt),\Phi(\theta_{0},t+dt))}{\Psi(\Theta(\theta_{0},t),\Phi(\theta_{0},t))}\notag\\
&=i\int_{0}^{2\pi}dt\bra{\Psi(\Theta(\theta_{0},t),\Phi(\theta_{0},t))}\partial_{t}\ket{\Psi(\Theta(\theta_{0},t),\Phi(\theta_{0},t))}\notag\\
&=-\mathrm{\mathrm{Im}}\int_{0}^{\theta_{0}}\int_{0}^{2\pi}d\theta dt\left(\partial_{\theta}\bra{\Psi(\Theta,\Phi)}\partial_{t}\ket{\Psi(\Theta,\Phi)}-\partial_{t}\bra{\Psi(\Theta,\Phi)}\partial_{\theta}\ket{\Psi(\Theta,\Phi)}\right)\notag\\
&=\int_{0}^{\theta_{0}}\int_{0}^{2\pi}d\theta dt\thinspace\tilde{B}(\theta,t),\label{Eq21}
\end{align}
			\end{figure*}
			
			where $\tilde{B}(\theta,t)$ is the Berry curvature introduced in
			Eq.~(\ref{eq:Berry_curvature}). Alternatively, using the mapping
			$\mathcal{F}:(\theta,t)\mapsto(\Theta,\Phi)$, the geometric phase
			can be expressed in terms of a curvature on the Bloch sphere as
			\begin{align}
			\chi_{\mathrm{geom}}(\theta_{0})-\chi_{\mathrm{geom}}(0)&=\int_{0}^{\theta_{0}}\int_{0}^{2\pi}d\theta dt\thinspace\tilde{B}(\theta,t)\notag \\ &=\int_{0}^{\theta_{0}}\int_{0}^{2\pi}d\theta dt\thinspace\frac{\partial(\Theta,\Phi)}{\partial(\theta,t)}B(\Theta,\Phi),\label{eq:GP-Blocch}
			\end{align}
			with
			\begin{align}
B(\Theta,\Phi)  =&\frac{\partial(\theta,t)}{\partial(\Theta,\Phi)}\tilde{B}(\theta(\Theta,\Phi),t(\Theta,\phi))\notag \\
 =&-\mathrm{Im}\left(\partial_{\Theta}\bra{\Psi(\Theta,\Phi)}\partial_{\Phi}\ket{\Psi(\Theta,\Phi)}\right. \notag \\ &\quad \quad  \quad -\left. \partial_{\Phi}\bra{\Psi(\Theta,\Phi)}\partial_{\Theta}\ket{\Psi(\Theta,\Phi)} \right) \notag \\ =&-\frac{1}{2}\sin\Theta(\theta,t),\label{eq:B}
\end{align}
			and where
			\begin{equation}
			\frac{\partial(\Theta,\Phi)}{\partial(\theta,t)}=\frac{\partial\Theta}{\partial\theta}\frac{\partial\Phi}{\partial t}-\frac{\partial\Theta}{\partial t}\frac{\partial\Phi}{\partial\theta}
			\end{equation}
			is the Jacobian of $\mathcal{F}$.
			
			The r.h.s. of Eq.~(\ref{eq:GP-Blocch}) admits a simple interpretation:
			$\chi_{\mathrm{geom}}(\theta_{0})-\chi_{\mathrm{geom}}(0)=-\mathcal{A}_{\theta_{0}}/2$,
			where $\mathcal{A}_{\theta_{0}}$ is the oriented area of the Bloch
			sphere covered by the measurement-induced trajectories $\ket{\Psi(\Theta(\theta,t),\Phi(\theta,t))}$
			with $\theta\in[0,\theta_{0}]$ (here the orientation of each infinitesimal
			contribution is given by the sign of the Jacobian). In particular,
			for $\theta_{0}=\pi$, $\mathcal{A}_{\pi}=4\pi$ if the surface generated
			by all the trajectories wraps the Bloch sphere once, and $\mathcal{A}_{\pi}=0$
			if it does not wrap around the Bloch sphere, cf.~Fig.~\ref{fig:trajectories}.
			This provides the two possible values for the Chern number (\ref{eq:chern-1})
			$\mathcal{C}\in\{0,-1\}$. Formally, this can be proven by explicitly
			using the degree of the map $\mathcal{F}$ \citep{Ambrosio2000}.
			The degree of the map $\mathcal{F}$, $\deg\mathcal{F}$, is the number
			of points $(\theta,t)$ that map to a given point $(\Theta,\Phi)$
			(provided that $(\Theta,\Phi)$ is a regular point of $\mathcal{F}$)
			taking the orientation into account. The degree does not depend on
			the specific point $(\Theta,\Phi)$ and can be expressed as
			\begin{equation}
			\deg\mathcal{F}=\sum_{(\theta,t)\in\mathcal{F}^{-1}[(\Theta,\Phi)]}\sgn\frac{\partial(\Theta,\Phi)}{\partial(\theta,t)},
			\end{equation}
			where $\mathcal{F}^{-1}[(\Theta,\Phi)]$ is the set of points $(\theta,t)$
			that are mapped by $\mathcal{F}$ into $(\Theta,\Phi)$, and $\sgn$
			is the sign function. Considering the integral as the sum of infinitesimal
			contributions and grouping those by the image points $(\Theta,\Phi)$,
			one then shows that
			\begin{equation}
			\int_{0}^{\pi}\int_{0}^{2\pi}d\theta dt\thinspace\frac{\partial(\Theta,\Phi)}{\partial(\theta,t)}\sin\Theta(\theta,t)=4\pi\deg\mathcal{F}.
			\end{equation}
			
			This topological feature is reflected in the discrete value of $\chi_{{\rm geom}}(\pi/2)$,
			hence in the $\theta$-dependence of $\chi_{{\rm geom}}$. To show
			this, note that Eq. (\ref{postselectedresult}) is symmetric under
			complex conjugation supplemented by $\theta\rightarrow\pi-\theta$,
			i.e. $\chi_{\mathrm{geom}}(\pi-\theta)=-\chi_{\mathrm{geom}}(\theta)\:\mod2\pi$.
			Using the continuity of $\chi_{{\rm geom}}(\theta)$, we obtain $\chi_{{\rm geom}}(\pi-\theta)=-\chi_{{\rm geom}}(\theta)+2\chi_{{\rm geom}}(\pi/2)$,
			and hence (for $\theta=0$), $\chi_{{\rm geom}}(\pi/2)=(\chi_{{\rm geom}}(\pi)+\chi_{{\rm geom}}(0))/2=\chi_{{\rm geom}}(\pi)/2=\pi\mathcal{C}$.
			Therefore, we have $\chi_{{\rm geom}}(\pi/2)=\pi\mathcal{C}=-\pi$
			for $c>c_{\mathrm{crit}}$ and $\chi_{{\rm geom}}(\pi/2)=\pi\mathcal{C}=0$
			for $c<c_{\mathrm{crit}}$ as shown in Fig.~\ref{fig:theta_dependence}.
			
			\subsection*{Monte Carlo numerical simulations}
			
			The results for the averaged GP have been obtained using a Monte Carlo
			simulation of the sum over different measurement readouts $\left\{ r_{k}\right\} _{k=1,...,N-1}$.
			We simulated the sequences of measurement readouts taking their probabilities
			$P_{\left\{ r_{k},r_{N}=+\right\} }=\abs{\sp{\psi_{0}}{\tilde{\psi}_{N-1}}}^{2}=\abs{\bra{\psi_{0}}\mathcal{M}_{N-1}^{(r_{N-1})}\dots\mathcal{M}_{1}^{(r_{1})}\ket{\psi_{0}}}^{2}=\abs{\bra{\psi_{0}}M_{\eta}(\mathbf{n}_{N-1},r_{N-1})\dots M_{\eta}(\mathbf{n}_{1},r_{1})\ket{\psi_{0}}}^{2}$
			into account. Namely, the quasicontinuous trajectory was represented
			by $N=500$ measurements ($N-1=499$ weak measurements and one strong
			postselected measurement). For the $k$-th measurement, we calculated
			$\vert\psi_{k}'(r_{k})\rangle=M_{\eta}(\mathbf{n}_{k},r_{k})\ket{\psi_{k-1}}\propto\ket{\tilde{\psi}_{k}}$
			($\ket{\psi_{k}}$ is the normalization of $\ket{\tilde{\psi}_{k}}$,
			which in turn has been defined in Eq.~(\ref{eq:psif})) and randomly
			determined the measurement readout $r_{k}=+/-$ according to probabilities
			$p(r_{k})=\sp{\psi_{k}'(r_{k})}{\psi_{k}'(r_{k})}$. Then, for the
			selected $r_{k}$, the normalized state $\ket{\psi_{k}}=\ket{\psi_{k}'(r_{k})}/\sqrt{p(r_{k})}$
			was calculated; after which the next measurement was simulated. After
			simulating $N-1$ weak measurements, $z_{\left\{ r_{k}\right\} }=z(\mathrm{realization})=\left(\sp{\psi_{0}}{\psi_{N-1}}\right)^{2}=P_{\left\{ r_{N}=+\right\} }e^{2i\chi_{\mathrm{geom}}(\left\{ r_{k}\right\} )}$
			was determined. After repeating this simulation $N_{\mathrm{realizations}}$
			times, $e^{i2\bar{\chi}_{\mathrm{geom}}-\alpha}=\langle e^{2i\chi_{\mathrm{geom}}}\rangle_{\mathrm{realizations}}=N_{\mathrm{realizations}}^{-1}\sum_{\mathrm{\mathrm{realizations}}}z(\mathrm{realization})$
			was calculated. Fig.~\ref{fig:geom_phase_statistics} was obtained
			using $N_{\mathrm{realizations}}=4000$.

            \subsection*{Data availability}

            All the data regarding the postselected geometric phase have been produced by means of an analytical formula, \eqref{postselectedresult}. The distribution of the geometric phases in the absence of postselection and the averaged geometric phase (Fig.~\ref{fig:geom_phase_statistics}) have been plotted based on the data produced by Monte Carlo simulations according to the algorithm described above. The code implementing the simulations and the relevant data can be found at \href{https://github.com/KyryloSnizhko/top-geom-meas}{https://github.com/KyryloSnizhko/top-geom-meas}.
	}
	
	\showmatmethods{} 
	
	\acknow{V.G. thanks T. Holder for helpful discussions. V.G. acknowledges financial
		support by the Minerva foundation under the Short-Term Research Grant
		programme. K.S. and Y.G. acknowledge funding by the Deutsche Forschungsgemeinschaft
		(DFG, German Research Foundation) -- Projektnummer 277101999 --
		TRR 183 (project C01). A.R. acknowledges support from EPSRC via Grant
		No. EP/P010180/1.}
	
	\showacknow{} 
	
\bibliography{GeomPhaseTopTransitionPaper}

\clearpage{}
\onecolumn
\part*{Supplementary material}

In this supplementary material, we focus on the full distribution
of trajectories and we define and study the averaged geometric phase
(GP). Subsequently, we propose an experimental protocol for the detection
of such an averaged phase based on Mach-Zehnder interferometry. Throughout
this Supplementary material, we use the same notation introduced in
the manuscript.

\section{Averaged geometric phase and its topological transition}

The most sensible way to define the averaged geometric phase for the
full distribution of trajectories is by appealing to averaging physically
measurable observables. Below, we propose a possible experimental
setup for detecting the GP, which introduces a direct protocol for
averaging over numerous closed trajectories. The averaged geometric
phase $\bar{\chi}_{{\rm geom}}$ is then defined through
\begin{equation}
e^{2i\bar{\chi}_{\mathrm{geom}}-\alpha}:=\langle e^{2i\chi_{\mathrm{geom}}}\rangle_{\mathrm{realizations}}=\sum_{\{r_{k}\}}\left(\bra{\psi_{0}}\mathcal{M}_{N-1}^{(r_{N-1})}\dots\mathcal{M}_{1}^{(r_{1})}\ket{\psi_{0}}\right)^{2},\label{defaverage-1}
\end{equation}
where $\chi_{{\rm geom}}$ is the geometric phase associated with
each single trajectory as introduced in the paper, $e^{-\alpha}$
is a suppression factor representing the suppression of the visibility
of interference in the experimental setup (cf. Section \ref{sec:Detection}
below) and the sum extends over all possible measurement readouts
$\{r_{k}\}$. The suppression factor accounts for two effects: (i)
the probability of a successful postselection in the final strong
measurement that ensures that the state trajectory is closed, and
(ii) ``dephasing'' due to $\chi_{\mathrm{geom}}$ having a spread
of values for different trajectories. The behavior of the averaged
GP $\bar{\chi}_{{\rm geom}}$ as a function of the measurement strength
has been reported in the paper (cf. Fig. (3) therein).

\begin{figure}
\begin{centering}
\includegraphics[width=0.9\columnwidth]{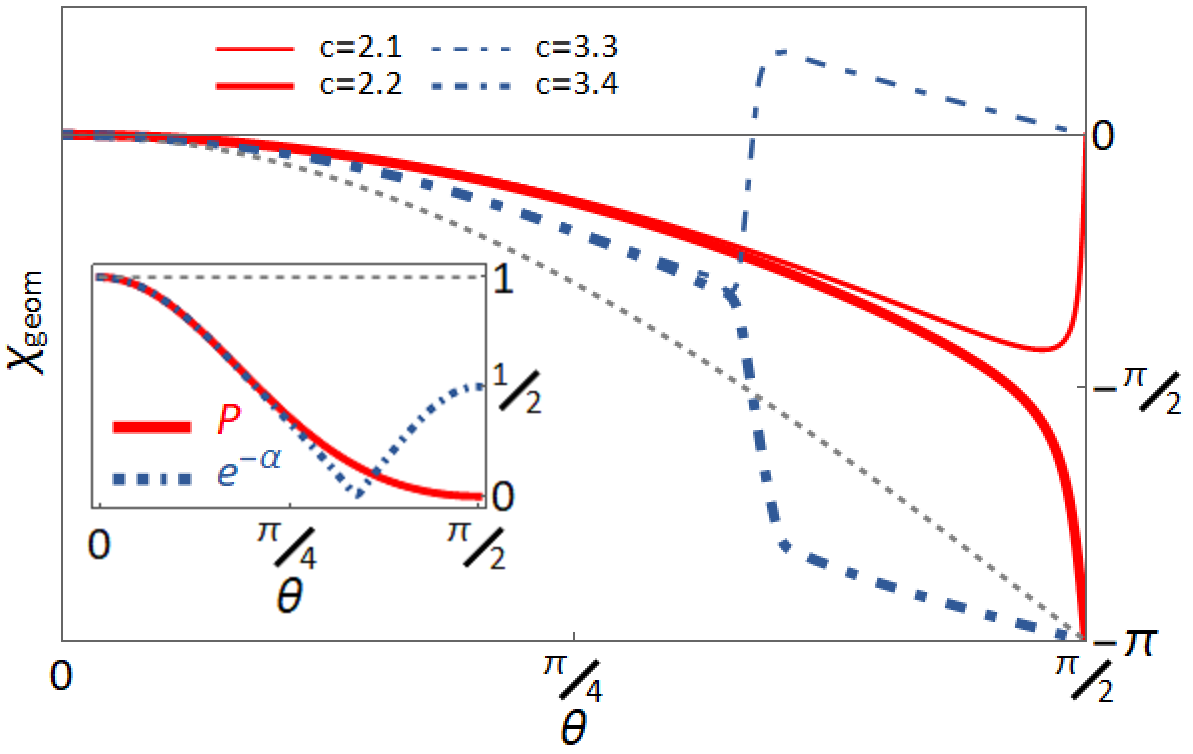}
\par\end{centering}
\caption{\textbf{Non-monotonicity and discontinuity of averaged geometric phases}.
We show the dependence of the averaged geometric phase on the polar
angle $\theta$ at different values of the integrated measurement
strength (cf. legend) for the averaged geometric phase (blue dot-dashed
lines) compared to that of the postselected geometric phase (red solid
lines). The ideal strong measurement dependence for $c\rightarrow\infty$
is presented as a grey dashed line. The dependence of the averaged
GP on $\theta$ displays an abrupt transition from monotonic to non-monotonic
behavior in the vicinity of $c=3.35$. The critical strength for the
averaged geometric phases differs from that of the postselected geometric
phase. The behavior is underlined by the fact that $\chi_{{\rm geom}}(\theta)$
can assume only discrete values, $0$ or $-\pi$, at $\theta=\pi/2$.
Inset: The suppression factor $e^{-\alpha}$ (blue dot-dashed line)
in the protocol with averaging at $c=3.3$ compared to the probability
of observing the most probable trajectory with postselected readout
sequence $\left\{ r_{k}=+\right\} $ (red solid line) at $c=2.1$;
the grey dashed line indicates $P=e^{-\alpha}=1$ for $c\rightarrow\infty$,
an asymptotic strong measurement. The plots for the protocol with
averaging have been obtained by Monte Carlo simulations with $N=500$
measurement steps per sequence and $N_{\mathrm{realizations}}=500000$
realizations\textbf{.\label{fig:theta_dependence-1}}}
\end{figure}

The dependence of the averaged GP $\bar{\chi}_{\mathrm{geom}}(\theta)$
on the polar angle $\theta$ of the measurement sequence presents
a transition as a function of the measurement strength in analogy
to the case of postselected measurement sequences (cf. Fig.~\ref{fig:theta_dependence-1}).
Yet, the features of the transition are different. We begin by noting
that $\bar{\chi}_{\mathrm{geom}}$ is defined only $\mod\pi$ (cf.
Eq.~(\ref{defaverage-1})) , i.e., $\bar{\chi}_{\mathrm{geom}}$
is defined on a circle $S^{1}$ of circumference $\pi$. This makes
a difference for the possible values of $\bar{\chi}_{\mathrm{geom}}$
at $\theta=\pi/2$: We have $\bar{\chi}_{\mathrm{geom}}(0)=0$ as
in the postelected case but at the equator $\theta=\pi/2$, where
the possible values of $\chi_{\mathrm{geom}}$ are only $0$ and $-\pi$,
both values correspond to $e^{2i\chi_{\mathrm{geom}}}=1$ implying
$\bar{\chi}_{\mathrm{geom}}(\theta=\pi/2)=0$. Hence, $\bar{\chi}_{\mathrm{geom}}(\theta)$
obeys periodic boundary conditions $\bar{\chi}_{\mathrm{geom}}(\pi/2)=\bar{\chi}_{\mathrm{geom}}(0)$,
allowing us to identify the points $\theta=0$ and $\theta=\pi/2$
such that $\theta$ can be defined on a circle $S^{1}$ of circumference
$\pi/2$. Therefore, $\bar{\chi}_{\mathrm{geom}}:S^{1}\ni\theta\mapsto\bar{\chi}_{\mathrm{geom}}\in S^{1}$
maps a circle onto a circle and can be classified by an integer-valued
winding number $m$ (how many times the function $\bar{\chi}_{\mathrm{geom}}(\theta)$
winds around the circle $S^{1}$ of length $\pi$ as $\theta$ varies
from $0$ to $\pi/2$). In the limit of infinitely weak measurements,
we have $\bar{\chi}_{\mathrm{geom}}(\theta)\equiv0$, yielding $m=0$.
In the limit of strong projective measurements, however, $\bar{\chi}_{{\rm geom}}(\theta)=\pi(\cos\theta-1)$,
yielding $m=-1$. If the function $\chi_{{\rm geom}}(\theta)$ would
depend continuously on the measurement strength, $m$ would be preserved
by increasing the measurement strength, which is incompatible with
the two limiting cases of strong and infinitely weak measurements.
Therefore, one expects a sharp transition at an intermediate measurement
strength $c=\bar{c}_{{\rm crit}}$, marking the jump between these
two different behaviors of $\bar{\chi}_{\mathrm{geom}}(\theta)$.
At this critical measurement strength, $\bar{\chi}_{\mathrm{geom}}(\theta)$
is ill-defined for a certain polar angle $\bar{\theta}_{\mathrm{crit}}$.
At these critical parameters ($\bar{\theta}_{\mathrm{crit}}\approx\pi/3$
with $c_{{\rm crit}}\approx3.35$), the visibility $e^{-\alpha}$
vanishes, cf.~Fig.~\ref{fig:theta_dependence-1} (inset).

With the $\left\{ r_{k}=+\right\} $ readout sequence being the most
probable, one naively expects the transition to take place near $c_{\mathrm{crit}}\approx2.15$.
However, precisely at the postselected transition ($c_{\mathrm{crit}}\approx2.15$,
$\theta_{\mathrm{crit}}=\pi/2$), the probability of this readout
sequence vanishes, rendering phase averaging over the remaining trajectories
a crucial factor. The actual transition happens at $\bar{c}_{{\rm crit}}\approx3.35$
and $\bar{\theta}_{\mathrm{crit}}\approx\pi/3$, when the contribution
of the $\left\{ r_{k}=+\right\} $ readout sequence is cancelled against
the phase-averaged contribution of the remaining sequences.

\section{Detection of the averaged geometric phase via Mach-Zehnder interferometry\label{sec:Detection}}

\begin{figure}
\centering{}\includegraphics[width=1\textwidth]{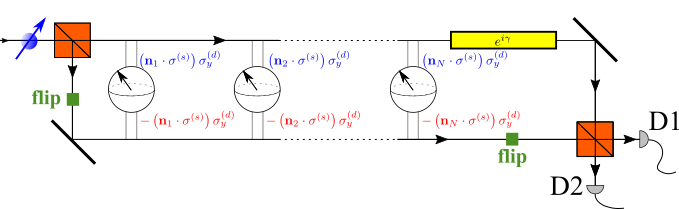}
\caption{\textbf{Experimental setups for observing the averaged measurement
induced GPs}. Scheme for observing the averaged GP, $\bar{\chi}_{\mathrm{geom}}$,
in a Mach-Zehnder interference setup. The detectors interact with
the particle in both interferometer arms according to different Hamiltonians,
$H_{{\bf n}}$ and $H_{-{\bf n}}$, cf.~Eq.~(\ref{eq:H-1}). Together
with the flip of the particle's internal degree of freedom in the
lower arm, this ensures that no ``which-path'' detection takes place,
and all readout sequences $\left\{ r_{k}\right\} $ contribute to
the interference pattern at D1 and D2. For any given readout sequence
$\left\{ r_{k}\right\} $, the GP accumulated by the particle is opposite
in the two arms of the interferometer. We assume an extra phase difference
$e^{i\gamma}$ produced by means other than measurements.}
\label{fig:interferometers-1}
\end{figure}

In order to observe the averaged measurement induced GP, we propose
an interferometric setup along the lines of the detection scheme described
for the detection of the postselected GP in the manuscript. Here,
however, a different approach to coupling the detectors to the polarization
of the beam in the interferometer arms is needed. Indeed, we need
to account for all readout sequences $\left\{ r_{k}\right\} $. Given
the initial detector state $+$, a readout $r_{k}=-$ may serve as
a \textquotedbl which-path\textquotedbl{} detection, undermining the
interference (the readout $r_{k}=+$ used for the postselected trajectory
in the manuscript does not provide ``which-path'' information due
to the properties of the null-type measurement we use). The only way
to overcome this handicap is to couple each detector to the two interferometer
arms, making it impossible to deduce from the readout signal which
arm the particle went through. This is demonstrated in Fig.~\ref{fig:interferometers-1}.
The $k-$th detector couples to $\sigma{}_{\mathbf{n}_{k}}=\boldsymbol{\sigma}\cdot\mathbf{n}_{k}$
in the upper arm and to $\sigma{}_{-\mathbf{n}_{k}}=-\sigma{}_{\mathbf{n}_{k}}$
in the lower arm of the interferometer. In addition, the particle's
inner degree of freedom in the lower arm is flipped before and flipped
back after the sequence of measurements. As a result, for any given
readout sequence $\left\{ r_{k}\right\} $, the trajectory on the
Bloch sphere corresponding to the lower arm is exactly opposite to
that of the upper arm (i.e., it is inverted with respect to the origin).
It follows that the solid angle $\Omega$ subtended by the trajectories
and the geometric phase $\chi_{\mathrm{geom}}$ accumulated through
the upper and the lower arms have opposite signs but same magnitudes.
Moreover, the probabilities $P_{\left\{ r_{k}\right\} }$ for yielding
the specific readout sequence $\left\{ r_{k}\right\} $ are exactly
the same in the two arms. This measurement scheme is thus completely
devoid of \textquotedbl which-path\textquotedbl{} signals. Provided
that the $N$-th measurement is postselected to yield $r_{N}=+$ and
the runs with $r_{N}=-$ do not contribute to the readings at drains
D1 and D2, the resulting intensities are
\begin{equation}
I_{1,2}=\frac{I_{0}}{2}\left(\sum_{\{r_{k}\}}\abs{\bra{\psi_{0}}\mathcal{M}_{N-1}^{(r_{N-1})}\dots\mathcal{M}_{1}^{(r_{1})}\ket{\psi_{0}}}^{2}\pm\mathrm{Re}\sum_{\{r_{k}\}}e^{i\gamma}\left(\bra{\psi_{0}}\mathcal{M}_{N-1}^{(r_{N-1})}\dots\mathcal{M}_{1}^{(r_{1})}\ket{\psi_{0}}\right)^{2}\right),\label{eq:averaged_intensity}
\end{equation}
where the second term in the parentheses is the interference term
expressible as $\mathrm{Re\,}e^{i\gamma+2i\bar{\chi}_{\mathrm{geom}}-\alpha}$
(cf. Eq.~(\ref{defaverage-1})). We provide a formal derivation of
this result below. This \emph{Gedankenexperiment} to detect the averaged
GP could be implemented in a variety of systems, e.g., optical systems
with absorptive polarizers or quantum dot detectors in electronic
interferometers.

\subsection{Output intensity of the averaged-phase interferometric detector}

The observed intensity in the detection scheme presented in Fig.~\ref{fig:interferometers-1}
is given by Eq. (\ref{eq:averaged_intensity}). This result is obtained
by analyzing the evolution of the compound system-detector state across
the interferometer. The collective state of the particle and all the
detectors after the initial beam splitter of the interferometer is
$\ket{\Psi_{i}}=\vert\psi_{0}\rangle\otimes[\ket{a=1}+\ket{a=-1}]\otimes\ket{+...+}/\sqrt{2}$,
where $a=\pm1$ describe the particle being in the upper or lower
arm respectively, and $\ket{+...+}$ is the initial state of all the
detectors. A ``flip'' (cf. Fig. \ref{eq:averaged_intensity}) applied
at the beginning and at the end of the lower arm acts on the system
via $R^{-1}({\bf n}_{0})\sigma_{x}^{(s)}R({\bf n}_{0})$. The interaction
of the system with each of the detectors is described by the Hamiltonian

\begin{equation}
\tilde{H}_{{\bf n}_{k}}=\lambda(t)[1-(\mathbf{n}_{k}\cdot\sigma^{(s)})\sigma_{z}^{(a)}]\sigma_{y}^{(d)}/2,\label{eq:H-1}
\end{equation}
where the Pauli $z$ matrix $\sigma_{z}^{(a)}$ acts on the degree
of freedom describing the occupation of the upper and lower arms of
the interferometer. The role of $\tilde{H}_{{\bf n}_{k}}$ is to let
the detector interact simultaneously with the upper and lower arms
via $H_{{\bf n}_{k}}$ and $H_{-{\bf n}_{k}}$, respectively, such
that the occupation of one of the two arms is not detected and ensuring
that the system in the upper and the lower arms accumulate opposite
geometric phases. After the interaction with all the detectors and
the action of the final flip (but before the particle passing through
the last beam splitter) the global state of the system and detectors
reads $\ket{\Psi_{f}}=[\vert\psi_{1}\rangle\ket{a=1}+\vert\psi_{-1}\rangle\ket{a=-1}]/\sqrt{2}$,
where $\vert\psi_{1}\rangle=\sum_{\{r_{k}\}}\ket{\{r_{k}\}}\mathcal{M}_{N}^{(r_{N})}\dots\mathcal{M}_{1}^{(r_{1})}\ket{\psi_{0}}e^{i\gamma}$
and $\vert\psi_{-1}\rangle=\sum_{\{r_{k}\}}\ket{\{r_{k}\}}R^{-1}({\bf n}_{0})\sigma_{x}^{(s)}R({\bf n}_{0})\mathcal{\tilde{M}}_{N}^{(r_{N})}\dots\mathcal{\tilde{M}}_{1}^{(r_{1})}R^{-1}({\bf n}_{0})\sigma_{x}^{(s)}R({\bf n}_{0})\ket{\psi_{0}}$,
where $\ket{\{r_{k}\}}$ is the state of the collection of detectors
the particle interacted with determined by the readout sequence $\{r_{k}\}$,
$\gamma$ is an extra phase that controls the interference pattern,
and the Kraus operators $\mathcal{M}_{k}^{(r_{k})}=M_{\eta_{k}}(\mathbf{n}_{k},r_{k})=R^{-1}(\mathbf{n}_{k})M_{\eta}(\mathbf{e}_{z},r_{k})R(\mathbf{n}_{k})$
and $\mathcal{\tilde{M}}_{k}^{(r_{k})}=M_{\eta_{k}}(-\mathbf{n}_{k},r_{k})=R^{-1}(\mathbf{n}_{k})\sigma_{x}^{(s)}M_{\eta}(\mathbf{e}_{z},r_{k})\sigma_{x}^{(s)}R(\mathbf{n}_{k})$.
The intensity of the output signals at $D_{1}$ and $D_{2}$ are
\[
I_{1,2}=I_{0}\bra{\Psi_{f}}(1\pm\sigma_{x}^{(a)})\ket{\Psi_{f}}/2.
\]
We now employ the fact that the last measurement is projective and
postselected to $r_{N}=+$, with the $r_{N}=-$ readout not taken
into account in calculating $I_{1,2}$. Therefore, $\mathcal{M}_{N}^{(r_{N})}=\mathcal{P}_{0}=\ket{\psi_{0}}\bra{\psi_{0}}=\ket{\mathbf{n}_{0}}\bra{\mathbf{n}_{0}}$
and $\mathcal{\tilde{M}}_{N}^{(r_{N})}=\ket{-\mathbf{n}_{0}}\bra{-\mathbf{n}_{0}}=R^{-1}({\bf n}_{0})\sigma_{x}^{(s)}R({\bf n}_{0})\ket{\mathbf{n}_{0}}\bra{\mathbf{n}_{0}}R^{-1}({\bf n}_{0})\sigma_{x}^{(s)}R({\bf n}_{0})$,
leading to
\begin{align}
\vert\psi_{1}\rangle= & \sum_{\{r_{k}\}}\ket{\{r_{k}\}}\ket{\psi_{0}}\times\left(\bra{\psi_{0}}\mathcal{M}_{N-1}^{(r_{N-1})}\dots\mathcal{M}_{1}^{(r_{1})}\ket{\psi_{0}}e^{i\gamma}\right),\\
\vert\psi_{-1}\rangle= & \vert\sum_{\{r_{k}\}}\ket{\{r_{k}\}}\ket{\psi_{0}}\times\left(\bra{\psi_{0}}R^{-1}({\bf n}_{0})\sigma_{x}^{(s)}R({\bf n}_{0})\mathcal{\tilde{M}}_{N-1}^{(r_{N-1})}\dots\mathcal{\tilde{M}}_{1}^{(r_{1})}R^{-1}({\bf n}_{0})\sigma_{x}^{(s)}R({\bf n}_{0})\ket{\psi_{0}}\right).
\end{align}
In order to simplify the expression for $\vert\psi_{-1}\rangle$,
we use the property
\begin{equation}
\bra{\psi_{0}}R^{-1}({\bf n}_{0})\sigma_{x}^{(s)}R({\bf n}_{0})\mathcal{\tilde{M}}_{N-1}^{(r_{N-1})}\dots\mathcal{\tilde{M}}_{1}^{(r_{1})}R^{-1}({\bf n}_{0})\sigma_{x}^{(s)}R({\bf n}_{0})\ket{\psi_{0}}=\left(\bra{\psi_{0}}\mathcal{M}_{N-1}^{(r_{N-1})}\dots\mathcal{M}_{1}^{(r_{1})}\ket{\psi_{0}}\right)^{*},\label{eq:oppositeness_of_phases}
\end{equation}
which we prove hereafter. Then, one immediately arrives at Eq.~(\ref{eq:averaged_intensity})
for $I_{1,2}$.

\paragraph{Computation of the phase accumulated through the lower arm.}

The evolution of the state through the lower arm entering the intensities
at the interferometer drain is computed via the property in Eq.~(\ref{eq:oppositeness_of_phases}),
which we prove here. We recall from the Methods in the manuscript
(cf.~Eqs.~(19,20) therein) that
\begin{equation}
\delta R=R(\mathbf{n}_{k+1})R^{-1}(\mathbf{n}_{k})=\begin{pmatrix}\cos^{2}\frac{\theta}{2}+e^{-2\pi i/N}\sin^{2}\frac{\theta}{2} & \frac{1}{2}(1-e^{-2\pi i/N})\sin\theta\\
\frac{1}{2}(1-e^{-2\pi i/N})\sin\theta & \sin^{2}\frac{\theta}{2}+e^{-2\pi i/N}\cos^{2}\frac{\theta}{2}
\end{pmatrix},\label{eq:again}
\end{equation}
and
\begin{equation}
M_{\eta}(\mathbf{e}_{z},+)=\left(\begin{matrix}1 & 0\\
0 & \sqrt{1-\eta}
\end{matrix}\right),\,\,M_{\eta}(\mathbf{e}_{z},-)=\left(\begin{matrix}0 & 0\\
0 & \sqrt{\eta}
\end{matrix}\right).\label{eq:again2}
\end{equation}
Using the hermiticity of $M_{\eta}(\mathbf{e}_{z},r_{k})$, and the
identities
\begin{equation}
\sigma_{x}^{(s)}\delta R\sigma_{x}^{(s)}=e^{-2\pi i/N}\sigma_{z}^{(s)}\delta R^{-1}\sigma_{z}^{(s)}=e^{-2\pi i/N}\sigma_{z}^{(s)}\delta R^{\dagger}\sigma_{z}^{(s)},
\end{equation}
\begin{equation}
\sigma_{z}^{(s)}M_{\eta}(\mathbf{e}_{z},r_{k})\sigma_{z}^{(s)}=M_{\eta}(\mathbf{e}_{z},r_{k}),
\end{equation}
for $\delta R$, we can write
\begin{align}
\bra{\psi_{0}}R^{-1}({\bf n}_{0})\sigma_{x}^{(s)}R({\bf n}_{0}) & \mathcal{\tilde{M}}_{N-1}^{(r_{N-1})}\dots\mathcal{\tilde{M}}_{1}^{(r_{1})}R^{-1}({\bf n}_{0})\sigma_{x}^{(s)}R({\bf n}_{0})\ket{\psi_{0}}\nonumber \\
 & =\bra{\mathbf{e}_{z}}\sigma_{x}^{(s)}\delta R\sigma_{x}^{(s)}M_{\eta}(\mathbf{e}_{z},r_{N-1})\sigma_{x}^{(s)}\delta R\sigma_{x}^{(s)}M_{\eta}(\mathbf{e}_{z},r_{N-2})\sigma_{x}^{(s)}\delta R...\delta R\sigma_{x}^{(s)}M_{\eta}(\mathbf{e}_{z},r_{1})\sigma_{x}^{(s)}\delta R\sigma_{x}^{(s)}\ket{\mathbf{e}_{z}}\nonumber \\
 & =e^{-2\pi iN/N}\bra{\mathbf{e}_{z}}\sigma_{z}^{(s)}\delta R^{\dagger}\sigma_{z}^{(s)}M_{\eta}(\mathbf{e}_{z},r_{N-1})\sigma_{z}^{(s)}\delta R^{\dagger}...\delta R^{\dagger}\sigma_{z}^{(s)}M_{\eta}(\mathbf{e}_{z},r_{1})\sigma_{z}^{(s)}\delta R^{\dagger}\sigma_{z}^{(s)}\ket{\mathbf{e}_{z}}\nonumber \\
 & =\bra{\mathbf{e}_{z}}\delta R^{\dagger}M_{\eta}(\mathbf{e}_{z},r_{N-1})\delta R^{\dagger}M_{\eta}(\mathbf{e}_{z},r_{N-2})\delta R^{\dagger}...\delta R^{\dagger}M_{\eta}(\mathbf{e}_{z},r_{1})\delta R^{\dagger}\ket{\mathbf{e}_{z}}\nonumber \\
 & =\left(\bra{\mathbf{e}_{z}}\delta RM_{\eta}(\mathbf{e}_{z},r_{1})\delta RM_{\eta}(\mathbf{e}_{z},r_{2})\delta R...\delta RM_{\eta}(\mathbf{e}_{z},r_{N-1})\delta R\ket{\mathbf{e}_{z}}\right)^{*}.
\end{align}
Using the explicit representation of
\begin{equation}
\ket{\mathbf{e}_{z}}=\begin{pmatrix}1\\
0
\end{pmatrix},
\end{equation}
(cf.~Eq.~(14) in the manuscript), we consider
\begin{align}
\bra{\mathbf{e}_{z}}\delta RM_{\eta}(\mathbf{e}_{z},r_{1})\delta RM_{\eta}(\mathbf{e}_{z},r_{2}) & \delta R...\delta RM_{\eta}(\mathbf{e}_{z},r_{N-1})\delta R\ket{\mathbf{e}_{z}}\nonumber \\
 & =\left(\bra{\mathbf{e}_{z}}\delta RM_{\eta}(\mathbf{e}_{z},r_{1})\delta RM_{\eta}(\mathbf{e}_{z},r_{2})\delta R...\delta RM_{\eta}(\mathbf{e}_{z},r_{N-1})\delta R\ket{\mathbf{e}_{z}}\right)^{T}\nonumber \\
 & =\bra{\mathbf{e}_{z}}\delta RM_{\eta}(\mathbf{e}_{z},r_{N-1})\delta RM_{\eta}(\mathbf{e}_{z},r_{N-2})\delta R...\delta RM_{\eta}(\mathbf{e}_{z},r_{1})\delta R\ket{\mathbf{e}_{z}},
\end{align}
where $T$ denotes transposition, and in the last step we used $\delta R^{T}=\delta R$
and $M_{\eta}(\mathbf{e}_{z},r)^{T}=M_{\eta}(\mathbf{e}_{z},r)$,
cf.~(\ref{eq:again}, \ref{eq:again2}). Finally, noticing that
\begin{equation}
\bra{\mathbf{e}_{z}}\delta RM_{\eta}(\mathbf{e}_{z},r_{N-1})\delta RM_{\eta}(\mathbf{e}_{z},r_{N-2})\delta R...\delta RM_{\eta}(\mathbf{e}_{z},r_{1})\delta R\ket{\mathbf{e}_{z}}=\bra{\psi_{0}}\mathcal{M}_{N-1}^{(r_{N-1})}\dots\mathcal{M}_{1}^{(r_{1})}\ket{\psi_{0}},
\end{equation}
one obtains Eq.~(\ref{eq:oppositeness_of_phases}), as desired.
	
\end{document}